\documentclass[a4paper,11pt]{article}
\pdfoutput=1
\usepackage[T1]{fontenc}
\usepackage[utf8]{inputenc}
\usepackage{jheppub}
\usepackage{float}
\usepackage{hyperref}
\usepackage{graphicx}
\usepackage{rotating}
\usepackage{tikz}
\usepackage{braket}
\usepackage{amsmath}
\usepackage{amssymb}
\usepackage{lmodern}
\usepackage{enumerate}
\usepackage{caption}
\usepackage{subcaption}
\usepackage[nottoc]{tocbibind}
\usepackage{accents}
\usepackage{trimclip}
\usepackage{dutchcal}
\usepackage{array} 
\usepackage{pbox}
\usepackage[makeroom]{cancel}
\usepackage[titletoc]{appendix}

\hypersetup{
    colorlinks=true,
    linkcolor=blue,
    citecolor=green,
    filecolor=magenta,      
    urlcolor=blue,
    linktocpage=true, 
    pdftitle={Overleaf Example},
    pdfpagemode=UseNone,
    }

\makeatletter
\newcommand{\thickhline}{%
    \noalign {\ifnum 0=`}\fi \hrule height 1pt
    \futurelet \reserved@a \@xhline
}
\newcolumntype{"}{@{\hskip\tabcolsep\vrule width 1pt\hskip\tabcolsep}}
\makeatother

\DeclareRobustCommand*{\phat}[1]{{\accentset{(\!\trimbox{0pt 1.1ex}{\ensuremath{\string^}}\!)}{#1}}}

\title{Electroweak symmetry non-restoration and suppressed dark radiation in  Supersymmetric Twin Higgs model
}

\title{Electroweak symmetry non-restoration and suppressed dark radiation in  Supersymmetric Twin Higgs model}
\author[1]{Marcin Badziak,}
\author[2,3,4]{Keisuke Harigaya,}
\author[1]{Ignacy Nałęcz}
\affiliation[1]{Institute of Theoretical Physics, Faculty of Physics, University of Warsaw, ul. Pasteura 5,\\
PL-02-093 Warsaw, Poland}
\affiliation[2]{Department of Physics, University of Chicago, Chicago, IL 60637, USA}
\affiliation[3]{Enrico Fermi Institute, Leinweber Institute for Theoretical Physics, and Kavli Institute
for Cosmological Physics, University of Chicago, Chicago, IL 60637, USA}
\affiliation[4]{Kavli Institute for the Physics and Mathematics of the Universe (WPI), The University of
Tokyo Institutes for Advanced Study, The University of Tokyo, Kashiwa, Chiba 277-8583,
Japan}

\emailAdd{inalecz@fuw.edu.pl}
\emailAdd{kharigaya@uchicago.edu}
\emailAdd{mbadziak@fuw.edu.pl}

\abstract{
We investigate a possibility of electroweak symmetry non-restoration (SNR) below the Twin electroweak scale ($\sim$TeV) within the Twin Higgs model. We focus on supersymmetric extensions with light sfermions where SNR is driven by mirror symmetry breaking in the Yukawa couplings. The inclusion of light scalars not only stabilizes the electroweak scale, but also extends SNR into new regions of the parameter space and enables a first-order phase transition. When this model is augmented with right-handed neutrinos with unbroken $B'-L'$ in the twin sector, the number of dark relativistic degrees of freedom can be reduced to the level consistent with the constraints from CMB data. The SNR in the supersymmetric Twin Higgs framework can naturally be integrated with minimal axiogenesis, offering a simultaneous explanation for the origin of baryon asymmetry and dark matter and the resolution of the strong CP problem that is consistent with astrophysical constraints.
}

\begin{document}

\maketitle
\flushbottom

\section{Introduction}

Twin Higgs (TH)~\cite{Chacko:2005pe,Chacko:2005vw,Chacko:2005un} is a class of models with a twin sector that couples to the Standard Model (SM) particles through the Higgs portal. It was introduced to alleviate the tension between the null result of LHC searches for new particles and the SM extensions such as supersymmetry (SUSY)~\cite{Falkowski:2006qq,Chang:2006ra,Craig:2013fga,Katz:2016wtw,badziak2017S,badziak2017M,badziak2017A} or Composite Higgs~\cite{Batra:2008jy,Barbieri:2015lqa,Low:2015nqa,Csaki:2015gfd,Contino:2017moj} that naturally explain the electroweak (EW) scale but postulate the existence of sub-TeV colored states. In this framework, the SM-like Higgs boson is a pseudo-Nambu-Goldstone boson (PNGB) of an approximate global symmetry group such as SU(4) or SO(8) which protects the Higgs mass from large radiative corrections from SM particles. This approximate global symmetry group is explicitly broken to $\mathbf{G}\times\mathbf{G}$, where $\mathbf{G}= \text{SU}(2)\times \text{U}(1)$ is the electroweak gauge group for both the SM and twin sector (TS). This paper focuses on the mirror TH, where an approximate mirror $\mathbb{Z}_2$ symmetry between the two sectors dictates the particle content and reduces the parameter space, alongside improving the radiative stability of the SM Higgs mass. For discussion of TH variant with a single-generation TS, see~\cite{Craig:2015pha}. 

TH currently evades stringent experimental constraints from the LEP and LHC~\cite{Barbieri:2005ri,Craig:2015pha,Katz:2016wtw,Ahmed:2017psb} in the significant part of its parameter space. However, the high-luminosity run of the LHC will further probe these models, and the remaining parameter space will be tested by the next generation of colliders~\cite{Dawson:2013bba, Asner:2013psa, TheATLAScollaboration:2014qxe, Buttazzo:2015bka}. Furthermore, the TS plays the role of a dark sector and particles from the TS are natural candidates for dark matter (DM)~\cite{Barbieri:2005ri,Craig:2015pha,GarciaGarcia:2015fol,Farina:2015uea,Craig:2015xla,Barbieri:2016zxn,Barbieri:2017opf,Badziak:2019zys,Terning:2019hgj,Chacko:2021vin,Badziak:2022eag,Holst:2023hff}, which can be probed in the forthcoming direct-detection experiments such as LUX-ZEPLIN or XENONnT~\cite{LZ:2018qzl,XENON:2020kmp}.

The structure of the TH scalar sector differs markedly from that of the SM, potentially altering the symmetry-breaking pattern in the thermal history of the Universe. Recent analysis of the scalar phase transition in TH revealed that it can be first-order without the need of introducing excessive tuning~\cite{Badziak:2022ltm}. This opens the way for explaining the observed excess of matter over antimatter in the Universe within the natural framework based on the electroweak-like baryogenesis in the TS. The main focus of the present paper is the other scalar phenomenon: EW symmetry non-restoration (SNR) below the $SU(4)$ breaking scale. We define it as the early breakdown of the SM $\text{SU}(2)\times \text{U}(1)$ symmetry followed by the freeze-out of the EW sphaleron process (the only known SM process that violates baryon and lepton numbers), which is not reversed at later stages of the Universe evolution. In the SM, the sphaleron process is in thermal equilibrium until around $T_\text{sph}\approx 130$ GeV~\cite{DOnofrio:2014rug}, which fixes the scale of any sphaleron-based mechanism that generates baryon asymmetry; any net baryon number generated by sphaleron processes at higher temperatures would be simply washed-out. Hence, the SNR is crucial for the models that utilize the EW sphaleron process to explain the relic baryon asymmetry but prefer somewhat larger scales. 

Non-restoration of gauge symmetry at finite temperature relies on a simple mechanism~\cite{Weinberg:1974hy}; positive thermal corrections to the scalar mass that drive this field to the symmetric vacuum are canceled by negative contributions from some other states. In the case of EW symmetry, positive thermal corrections come from the SM fields, while the negative contributions require new fields coupled to the SM Higgs. These new fields must be in thermal equilibrium at $T_\text{sph}$, implying their masses are roughly at the EW scale.%
\footnote{In some PNGB models which do not introduce new neutral states at the EW-scale, the SM EW gauge symmetry is broken at $T\gg T_\text{sph}$~\cite{Espinosa:2004pn, Aziz:2009hk, Ahriche:2010kh}. Nevertheless, such early symmetry breaking is followed by a short period, during which the symmetry is temporarily restored and the sphaleron process reenters thermal equilibrium before eventually being broken again at $T=0$. We emphasize that in these scenarios baryon asymmetry generated by EW sphaleron processes at any $T>T_{\text{sph}}$ would be later washed out.}
Given no new particles were found at this scale, the new states are expected to be SM singlets. Extensions of the SM featuring SNR include models with a large number of neutral scalars~\cite{Baldes:2018nel,Glioti:2018roy,Meade:2018saz, Carena:2021onl} or neutral fermions with substantial couplings to the Higgs boson~\cite{Matsedonskyi:2020mlz}. The SNR is also observed in the models where additional scalar fields coupled to the Higgs field develop non-zero VEVs~\cite{Glioti:2018roy,Co:2019wyp, Matsedonskyi:2021hti}.

TH models by construction introduce many new states that are coupled to the Higgs and transform as singlets under the SM gauge group. In the mirror TH, SNR can be driven by heavy twin fermions~\cite{Matsedonskyi:2020kuy,Badziak:2022ltm}, though this requires hard breaking of the $\mathbb{Z}_2$ symmetry to enhance TS quark and lepton Yukawas, leading to significant radiative contributions to the SM Higgs mass parameter. Weak collider bounds on the colorless fermion partners make it relatively easy to find a UV completion that balances these radiative corrections from asymmetric lepton Yukawas. In contrast, in UV-completions of the quark sector, the cutoff scale typically must be at least at a few TeV. Consequently, radiative contributions from the quark sector to the SM Higgs mass often persist, thereby spoiling the naturalness of the model.

In this paper, we demonstrate that both \emph{scalars and fermions} in the twin sector can contribute to SNR if the Yukawa couplings of twin fermions are enhanced and the scalar-Higgs couplings meet certain conditions. We consider a supersymmetric TH model which includes a twin copy of Minimal Supersymmetric Standard Model (MSSM) enhanced twin Yukawas of the charged leptons and their scalar light partners. Light twin scalar leptons cancel radiative contributions from the charged twin leptons to the SM Higgs mass and simultaneously enhance the SNR effect across a significant part of the parameter space. Moreover, they give rise to the barrier in the scalar potential, allowing for the combination of SNR with a first-order phase transition (FOPT), which does not occur when the SNR is driven only by fermions.\footnote{In ref.~\cite{Matsedonskyi:2022btb} a SUSY model was proposed in which EW SNR was present together with first-order electroweak transition. However, in contrast to a present paper the scalar degrees of freedom introduced to generate FOPT did not help to achieve SNR.} Nonetheless, $\mathbb{Z}_2$ breaking in the lepton sector alone is insufficient to induce the SNR; one must also significantly enhance the Yukawas of the twin quarks, which reintroduces some tuning.

We show that this additional tuning is significantly ameliorated in an extension of the model with three right-handed neutrinos (RHNs) and their light SUSY partners. This can be realized if both $B-L$ and  twin $B'-L'$ symmetries are gauged, but only the $B-L$ symmetry is broken. In the spirit of the type-I seesaw mechanism, we take neutrino Yukawas to be of order one and assume that a large SM Majorana mass term gives small SM neutrino masses, while in TS, unbroken $B'-L'$ symmetry forbids the Majorana mass terms. $\mathcal{O}(1)$ twin neutrino Yukawa couplings can also be obtained via inverse seesaw mechanism and we construct a SUSY UV completion of TH models that includes right-handed neutrinos. Twin neutrinos and their scalar partners with order one Yukawa couplings take over a substantial part of the $\mathbb{Z}_2$-breaking necessary for the SNR from the quark sector, which reduces the overall tuning of the model.

Enhancing TS fermion masses not only induces SNR but also mitigates the problem of excessive dark radiation, which is characteristic of the mirror TH. (For solutions based on asymmetric particle decays, see~\cite{Chacko:2016hvu,Craig:2016lyx,Curtin:2021alk}.) The problem arises because the numerous neutral states postulated in TH are in equilibrium with the thermal plasma and carry a substantial fraction of the Universe's entropy. When the two sectors decouple, this entropy is trapped in the TS radiation, which is inconsistent with precise observations of the cosmic microwave background (CMB)~\cite{Planck:2015fie, Planck:2018vyg} and the observed yields from primordial nucleosynthesis~\cite{Cyburt:2004yc, Cyburt:2015mya}. Interestingly, the natural realization of SNR in TH prefers twin quarks and leptons to be much heavier than the temperature at which the two sectors decouple, lowering the amount of dark radiation. By adjusting the mass of the lightest TS quark in the model with large twin lepton Yukawa couplings, the amount of dark radiation can be made consistent with primordial nucleosynthesis and marginally consistent with measurements of the cosmic microwave background, while the variant based on unbroken gauged $B'-L'$ symmetry with order one Yukawa couplings for twin RH neutrinos can be fully aligned with cosmological constraints~\cite{Barbieri:2016zxn}.

As an example of a particular baryogenesis model that could benefit from our construction, we consider minimal axiogenesis~\cite{Co:2019wyp} that utilizes a rotating QCD axion field to generate chiral asymmetry, which is then reprocessed by EW sphaleron processes to baryon abundance. This mechanism has the advantage of resolving the strong CP problem \emph{and} explaining the DM abundance and the baryon asymmetry of the Universe. In this scenario axions are produced by the kinetic misalignment mechanism~\cite{Co:2019jts} and the axion abundance fits the observed energy density of DM, together with the observed baryon asymmetry, for the axion decay constant $f_a$ in the range between $10^6$ and $10^7$~GeV in minimal axiogenesis. However, such small values of $f_a$ are excluded by many astrophysical constraints, in particular from SN1987A and neutron star cooling~\cite{Carenza:2019pxu,Buschmann:2021juv}. The problem could be resolved either by introducing astrophobic axions~\cite{DiLuzio:2017ogq,Bjorkeroth:2019jtx,Badziak:2023fsc,Badziak:2024szg} with very small couplings to nucleons, electrons and photons or by increasing the EW sphaleron decoupling temperature $T_{\text{sph}}$. We show that combining axiogenesis with TH featuring SNR achieves the latter option, addressing multiple SM issues within a single, relatively simple, self-consistent framework.

The paper is structured as follows. Section \ref{sec:LSR} introduces the TH scalar potential, the main tool for analyzing Higgs phase thermodynamics and discusses the mechanism of the SNR in the SM sector, listing conditions required to keep SM sphalerons decoupled at low temperatures. Section \ref{sec:SUSYTH} is devoted to a family of scenarios that feature SNR in the considerable part of their parameter space. It is also shown that these exemplary TH extensions inherently resolve the problem of dark radiation in TH. Section \ref{sec:SNRaxio} briefly reviews axiogenesis -a mechanism consistent with TH that naturally explains DM and relic baryon abundance, provided early breaking of the EW symmetry. Finally, section \ref{sec:SUM} summarizes our findings and comments on possible further applications of our results.

\section{Symmetry non-restoration}\label{sec:LSR}

\subsection{Twin Higgs scalar potential}\label{sec:Veff}

The scalar potential determines the evolution of the SM Higgs and its twin. In particular, it encodes the information about the scale of the SM EW symmetry breaking, and therefore investigation of it is the main objective in our discussion of SNR.

The tree-level part of the scalar potential is\footnote{Here, we assume that $\mathbb{Z}_2$ is softly broken in the scalar potential and ignore a possible $\mathbb{Z}_2$-breaking quartic term. For the extensive discussion of its impact on EWSB and naturalness see~\cite{Katz:2016wtw}, while the impact on the order of scalar PT is discussed in~\cite{Badziak:2022ltm}.}
\begin{equation}\label{eq:Vtree}
V_{\text{tree}}(H_A,H_B)=\lambda (|H_{A}|^2+ |H_{B}|^2-f_0^2/2)^2+\kappa (|H_{A}|^4+|H_{B}|^4)+f_0^2\sigma |H_{A}|^2,
\end{equation}
with $H_A$ and $H_B$ being the SM and TS scalar doublets, respectively. Each of them contains three would-be Nambu-Goldstone modes, and we can replace each doublet in \eqref{eq:Vtree} with the dynamical mode, $H_{A,B}\rightarrow h_{A,B}/\sqrt{2}$.

The tree-level part of the potential contains four free parameters $\lambda$, $\kappa$, $\sigma$, and $f_0$. $\lambda$ and $f_0$ are $\mathbb{Z}_2$ and $SU(4)$ invariant, $\kappa$ is $\mathbb{Z}_2$ invariant but $SU(4)$ breaking, and $\sigma$ is both $\mathbb{Z}_2$ and $SU(4)$ breaking.
In order to stabilize the SM Higgs mass by the Twin Higgs mechanism, $SU(4)$ symmetric coupling $\lambda$ has to dominate over the other couplings, i.e., $\lambda\gg \kappa,\; \sigma$. The TH scale $f_0$ is bounded from below by the Higgs couplings measurements and the constraints on the branching ratio of the Higgs boson to invisible particles at the LHC~\cite{ATLAS:2022vkf,CMS:2022dwd} which translate to the lower bound~\cite{Ahmed:2017psb} $f_0\gtrsim3\,v_{\text{EW}}$. On the other hand, the tuning between the EW and TH scales in mirror TH goes as $f_0^2/v^2_{\text{EW}}$~\cite{Craig:2013fga} so that $f_0$ cannot be too large. In our numerical results, with an exception in section \ref{sec:T_SNR}, we use reference values
\begin{equation}
    \lambda=1,\qquad  f_0=1\;\text{TeV}.
\end{equation}
The remaining two symmetry-breaking couplings are not subject to any experimental constraints.

In the PNGB limit $\lambda\gg \kappa,\;\sigma$ and $v_A \ll f_0$, the expressions for the tree-level VEVs and masses of the SM and twin Higgs bosons are\footnote{In this paper we work in the region where the PNGB approximation is accurate since this minimizes the tuning. The full expressions for the tree-level Higgs masses and VEVs can be found in~\cite{Ahmed:2017psb}.}
\begin{equation}\label{eq:VEV-s}
v_A^2\approx \frac{1}{2}f_0^2(1-\frac{\sigma}{\kappa}),\qquad v_B^2\approx f_0^2-v_A^2 \,,
\end{equation}
and
\begin{equation}\label{eq:masses}
m_h^2\approx 4\kappa v_A^2,\qquad m_{h'}^2\approx 2 f_0^2 \lambda,
\end{equation}
where $m_h$ denotes the SM-like Higgs boson mass and $m_{h'}$ is its twin counterpart. The conditions for the correct EW vacuum $v_A\approx 246\; \text{GeV}$ and SM Higgs mass $m_h\approx 125\; \text{GeV}$ fix two combinations out of four free parameters.%
\footnote{For all numerical results, we solved one-loop conditions for the SM Higgs mass and VEV at $T=0$ in order to eliminate $\kappa$ and $\sigma$.}
Using the relation, for given $\lambda$ and $f_0$, we determine symmetry breaking couplings $\kappa$ and $\sigma$.

To analyze the evolution of the Higgs field values at finite temperature, we also account for leading-order quantum and thermal corrections to the effective potential using the Parwani method~\cite{Parwani:1991gq},
\begin{equation}\label{eq:V_tot}
V_{\text{tot}}=V_{\text{tree}}(h_A, h_B)+V_{\text{CW}}(\bar{m}_i(h_A,h_B,T))+V_{\text{therm}}(\bar{m}_i(h_A,h_B,T),T)
\end{equation}
which, apart from the tree-level contribution \eqref{eq:Vtree}, contain the one-loop Coleman-Weinberg zero-temperature corrections and thermal contributions that depend on the masses of particles in the model. The explicit expressions for $V_{\text{CW}}$ and $V_{\text{therm}}$ can be found in appendix~\ref{app:1loop}.
The leading order effective potential with resummation is obtained by replacing the zero-temperature masses $m_i(h_A,h_B)$ in the one-loop expressions with thermally corrected ones $\bar{m}_i(h_A,h_B,T)$. 
For details, see appendix \ref{app:resummation}.

In order to evaluate the one-loop effective potential \eqref{eq:V_tot}, one needs to choose a gauge. Throughout this paper we work in the Landau gauge, which include the would-be Nambu-Goldstone modes in both SM and Twin sectors, but with the following approximation. Since the leading-order would-be Nambu-Goldstone contributions to $V_{\text{CW}}$ do not affect the position of the true vacuum~\cite{Martin:2014bca, Elias-Miro:2014pca} and hence have no impact on the SNR, we neglect them in our computations. To avoid imaginary contributions to $V_{\text{tot}}$, in the thermal potential $V_{\rm therm}$ we keep only the leading contribution from would-be Nambu-Goldstone bosons in the high-temperature expansion, while we use full thermal correction from all other states. While all the plots presented in this article are obtained using the above methodology, we have checked that performing computations in the unitary gauge leads to similar results.

\subsection{Scalar and fermion corrections to Higgs potential}\label{sec:SFcorr}

In the SM at $T=0$, the square of the Higgs mass at the SU(2)-preserving point is negative and the Higgs obtains a VEV that breaks the EW symmetry. On the other hand, at high temperatures, positive thermal corrections to the Higgs mass usually dominate over negative tree-level contributions and the EW symmetry is effectively restored. This mechanism is universal; in the SM and a great majority of its extensions, the EW symmetry restoration occurs roughly at $T\sim 100\;\text{GeV}$. Nevertheless, there exists a class of models where some thermal contributions to the SM Higgs mass are negative, and the EW symmetry restoration takes place at higher temperatures.

In this section we show that SNR in TH can be induced by scalars and fermions coupled to the twin Higgs since they give thermal corrections to the effective potential that keep the SM Higgs field away from the symmetric point. To illustrate how these corrections arise, let us compute scalar and fermionic thermal contributions to the effective potential in the effective field theory (EFT) with the radial mode integrated out, where $h_B$ and $h_A$ are related via
\begin{equation}\label{eq:EFT}
    h_B\approx \sqrt{f_0^2-h_A^2},
\end{equation}
which is accurate as long as $T\ll f_0$ and $T\ll m_{h'}$. We also expand the thermal potential in $m^2/T^2$, with $m$ being the mass of particles relevant for the phase transition, to the leading order in the case of fermions and to the next-to-leading order for bosons. These approximations are valid in the region with $m\ll T\ll f_0$. We assume here that for $T<m$ the EW symmetry remains broken since the thermal effects are dominated by the tree-level potential, which drives the minimum towards its zero temperature position. On the other hand, at high temperatures $T\sim f_0$, we expect that both EW and twin EW symmetries are restored. In the next sections we show that these two assumptions hold for all benchmarks points featuring SNR, and as a result, thermal VEVs evolution always follows the path similar to the one shown in figure~\ref{fig:Benchmark}.\footnote{The plots in figure \ref{fig:Benchmark} were obtained by explicit minimization of the effective potential \eqref{eq:V_tot} for a valid TH extension introduced in section~\ref{sec:SNRpheno}, without imposing eq. \eqref{eq:EFT} and no other assumptions about the scalar VEVs at low and high temperatures.}

Let us first compute the contribution to the effective potential \eqref{eq:V_tot} from the SM-like fermion species $i$ and its twin,
\begin{equation}\label{eq:dVferm}
\begin{aligned}
  \delta V_i (h_A,\sqrt{f_0^2-h_A^2})&=-\frac{n_i T^4}{2\pi^2}\left[J_F\left(\frac{m_i ^2}{T^2}\right)+J_F\left(\frac{\hat{m}_i ^2}{T^2}\right)\right]\\
  &=T^2\,h_A^2\frac{n_i(y^2_i- \hat{y}^2_{i}) }{96}+\mathcal{O}\Big(\frac{\phat{m}{}^4_i}{T^4}\Big)
\end{aligned}
\end{equation}
where $J_{B/F}(x)$ stand for thermal integrals defined in eq. \eqref{eq:JBJF}, $y_i$ is the fermion Yukawa while $n_i$ is the number of its degrees of freedom. Here and after, we use hat to denote twin masses and couplings and $\phat{\;}$ is used to denote masses and couplings in TS \emph{or} SM sector. In the above equation, we neglected the field-independent terms, as they do not affect the evolution of Higgs VEVs. If one enhances $\hat{y}_i$ w.r.t. $y_i$ by introducing hard breaking of $\mathbb{Z}_2$ symmetry, the contribution to the quadratic term of $h_A$ becomes negative and opposes the symmetry restoration. The rough condition for fermion-induced SNR in TH reads~\cite{Matsedonskyi:2020kuy}
\begin{equation}\label{eq:SNRF}
    \sum_{j\in\text{f.}}\frac{n_j}{4} (\hat{y}^2_j-y^2_j)\geq5,
\end{equation}
where the sum runs over all fermionic species in equilibrium with the thermal plasma.

In analogy to the fermionic case, we also compute the contribution to the effective potential from scalar species labeled by $k$ and its TS partner. For simplicity, we assume here that it does not obtain VEV for $T<f_0$ and its mass is $m_k ^2=\mu^2_0+\xi^2 h_A^2/2+\omega T^2$, where $\omega$ is the dimensionless coefficient of the daisy correction defined in app.~\ref{app:resummation}. The contribution to $V_{\text{eff}}$ in the EFT approximation and high $T$ limit reads
\begin{equation}\label{eq:dVscal}
\begin{aligned}
  &\delta V_{k}(h_A,\sqrt{f^2_0-h_A^2})=\frac{n_k T^4}{2\pi^2}\left[J_B(\frac{m_k ^2}{T^2})+J_B(\frac{\hat{m}_k ^2}{T^2})\right]\\ 
  &\approx n_k\,\frac{T^2}{48} h^2_A\left[\xi^2\left(1-\sqrt{\frac{\mu^2_0}{T^2}+\omega}\right) -\hat{\xi}^2\left(1-\sqrt{\frac{2\hat{\mu}^2_0+\hat{\xi}^2 f_0^2}{2T^2}+\hat{\omega}}\right)\right]+T^4\,\mathcal{O}\Big(\frac{\phat{m}{}^3_k}{T^3},\frac{h_A^3}{T^3}\Big),
\end{aligned}
\end{equation}
where $n_k$ is the number of degrees of freedom of scalar $k$.

It follows from the above equation that scalars can also provide contributions which oppose the symmetry restoration, although this is possible only for specific scalar coupling values. On one hand, $\hat{\xi}$ has to be larger than $\xi$ to dominate over the SM contribution. On the other hand, beyond a certain threshold, the $\hat{\xi}$-dependent term changes  sign and the scalar contribution to $V_{\text{eff}}$ does not favor SNR anymore. In the limit $\hat{\xi}\gg\xi$, the threshold value is
\begin{equation}\label{eq:maxxi}
    \hat{\xi}_{\text{tr}}=\frac{T_{\text{min}}}{f_0}\,\sqrt{2\Big(1-\hat{\omega}-\frac{\hat{\mu}^2_0}{T_{\text{min}}^2}\Big)},
\end{equation}
with $T_{\text{min}}\sim100\;\text{GeV}$ being the temperature at which tree-level effects start to dominate the thermal evolution of the SM Higgs VEV. Generically, the threshold \eqref{eq:maxxi} is low and thus one needs to introduce large number of scalar degrees of freedom to obtain purely scalar SNR. However, $\hat{\xi}_{\text{tr}}$ can be significantly increased if one considers a model where $\mu^2_0$ is negative.

Finally, we consider a model where SNR is as a result of a combined effect of the scalar and fermionic contributions. Since the scalar- and fermion-induced terms in $V_{\text{tot}}$ are independent from each other, the condition for SNR is simply
\begin{equation}\label{eq:SNR}
   \frac{12}{T_{\text{min}}^2 }\frac{\partial^2\;}{\partial h_A ^2}\Big\rvert_{T=T_{\text{min}}}\Big( \sum_{k\in\text{s.}}\,\delta V_k +\sum_{i\in\text{f.}}\delta V_i\Big)\geq5,
\end{equation}
where the first sum runs over all the scalars and the second over all fermions in the model, while $\delta V_i$ and $\delta V_k$ are defined in \eqref{eq:dVferm} and \eqref{eq:dVscal}.

It is usually problematic to construct a model which incorporates sufficient amount of hard $\mathbb{Z}_2$ breaking to satisfy the condition \eqref{eq:SNRF} for the fermionic SNR without excessive tuning (see however~\cite{Matsedonskyi:2020kuy}). On the other hand, if the early symmetry restoration is caused solely by scalars, the number of postulated scalar states has to be large, which renders the model non-minimal. In contrast, the SNR induced by fermions and scalars can be easily accommodated in a vast class of SUSY TH models. In section~\ref{sec:SUSYTH}, we propose SUSY extensions of TH which allows for SNR due to the combined effect of scalar and fermion thermal contributions. The proposed model does not introduce a large number of exotic scalars and does not suffer from large tuning from hard $\mathbb{Z}_2$ breaking in fermion Yukawa couplings.

\subsection{Symmetry non-restoration and phase transitions}\label{sec:FOPT}

There are also interesting consequences of explicit $\mathbb{Z}_2$-breaking in the fermion-Higgs and scalar-Higgs couplings for cosmological phase transitions. These events often lead to highly non-equilibrium dynamics and therefore are of great phenomenological interest. They play a crucial role in many scenarios of matter-antimatter symmetry, can potentially lead to the emission of the stochastic gravitational-wave background~\cite{Caprini:2015zlo,Caprini:2019egz} or the production of primordial black holes~\cite{Hawking:1982ga, Kodama:1982sf}. Even though all phase transitions in the SM are expected to be smooth, explicit breaking of $\mathbb{Z}_2$ symmetry between the Twin and SM sectors can be used to make two of the transitions in this model first-order. One is an EW-like phase transition which takes place roughly at $100$ GeV scale~\cite{Badziak:2022ltm} and another is a twin QCD quark confinement phase transition which occurs typically at $\mathcal{O}(0.1-1)$ GeV above the SM QCD phase transition temperature. In this section, we analyze the order of the EW-like transition and comment on the twin QCD phase transition in TH variants that feature SNR. 

The EW-like phase transition takes place when the Higgs fields tunnel from a supercooled metastable vacuum state of the effective potential \eqref{eq:V_tot} to the true vacuum. The crucial feature of this transition is the height of the potential barrier separating the true and false vacua. Only if the barrier is sufficiently high, the transition is of first-order. Estimation of the potential barrier height in models where more than one scalar field undergoes phase transition usually cannot be approached analytically. Generally, one first needs to find the tunneling path in the scalar field space, which requires solving a system of differential equations. Luckily, whenever the scalar transition in TH is combined with SNR, the Higgs VEV evolution closely follows the simple EFT trajectory \eqref{eq:EFT} in the $h_A-h_B$ space. Along this trajectory we can parameterize the dynamical scalar fields as
\begin{equation}
    h_A^2=f^2_0\, x,\qquad\qquad  h_B^2=f^2_0\,(1-x),
\end{equation}
with $x\in[0,1]$. 

The EW-like FOPT usually occurs at temperatures at which the potential is dominated by the thermal part. The thermal contributions from the single bosonic/fermionic species in the high-temperature regime read\footnote{For clarity, we compute only the contribution to the barrier from SM particles. The results would hold also for the TS particles since the barrier height is not affected by the reparametrization $x\rightarrow 1-x$.}
\begin{align}
    &\delta V_F(x)=T^4\Big[\frac{n_i}{96} \frac{f_0^2}{T^2} y^2_i \, x+\,\mathcal{O}(\frac{m_i^4}{T^4})\Big],\\
    &\delta V_B(x)=T^4\Big[\frac{n_i}{96} \frac{f_0^2}{T^2} g^2_i\, x-\frac{n_i}{96\pi} \frac{f_0^3}{T^3} g_i^3\, x^{\frac{3}{2}}+\mathcal{O}(\frac{m_i^4}{T^4})\Big].
\end{align}
Clearly, at the leading order in $m_i^2/T^2$, the thermal corrections are linear in $x$ and thus, neither fermionic nor bosonic terms contribute to the barrier. The first non-linear term $x^{3/2}$ appears at the order $m_i^3/T^3$ from $\delta V_B(x)$. The resulting function
\begin{equation*}
    \delta V_F(x)+\delta V_B(x)=\gamma x - \zeta x^\frac{3}{2}+\mathcal{O}(\tfrac{f_0^4}{T^4})
\end{equation*}
has a single maximum separating two minima as long as $\gamma\sim\zeta$, and thus provides the main contribution to the potential barrier.\footnote{If the barrier is roughly symmetric under $x\rightarrow 1-x$, the contributions from the twin sector will add up to the SM contributions and amplify the barrier. Otherwise, the barrier will be M-shaped, with the third intermediate minimum.} Therefore, sufficient number of bosonic degrees of freedom is necessary to rise $\zeta$ and get a potential barrier along the tunneling path. 

It follows that TH scalar transition is of the first order if and only if the number of scalars and gauge bosons coupled to Higgs fields is sufficiently large compared to fermionic degrees of freedom, since only in this case the condition $\gamma\sim\zeta$ may be fulfilled. A direct consequence of this observation is that \textbf{in TH or any other PNGB models SNR obtained by enhancing only twin fermion couplings excludes the possibility of EW-like FOPT} since a large number of strongly coupled twin fermions sources large $\gamma$ while leaving $\zeta$ untouched. Nevertheless, scalar FOPT and SNR can be combined in TH, provided a sufficient number of strongly coupled bosons in the TS. In section~\ref{sec:SUSYTH} we illustrate this statement with a simple SUSY TH extension. This model can feature SNR caused by fermions only as well as by the combined effect of fermions and supersymmetric scalars. We numerically find that the strong FOPT of the Higgs fields is possible only in the latter case.

Let us also briefly comment on an impact of our scenario on 
the twin QCD phase transition, which occurs when quark-gluon plasma confines into hadrons. At the confinement scale, perturbative computation techniques cease to work and one needs to rely on lattice results. The order of the QCD phase transition crucially depends on the quark masses. In the SM, the QCD confinement phase transition is known to be smooth~\cite{Aoki:2006br,Bhattacharya:2014ara,Aarts:2023vsf}. However, in the TS, fermion Yukawas may be enhanced by introducing $\mathbb{Z}_2$ breaking, leaving no nearly massless quarks. In such a scenario, twin QCD phase transition becomes first-order~\cite{Svetitsky:1982gs,Panero:2009tv,Petreczky:2012rq}. 

Being a first-order transition, confinement in the twin QCD may produce a stochastic gravitational-wave background. Estimation of the resulting experimental signal requires the knowledge of the transition temperature, latent heat, and its inverse duration time. While there are robust computations of the first two parameters~\cite{Sommer:1993ce, Caselle:2018kap}, lattice results for the transition duration are not available, but most of analytical estimates predict that it is too large to yield any detectable signals.\footnote{On one hand, computations based on the Polyakov Loop approximation~\cite{Halverson:2020xpg,Huang:2020crf, Kang:2021epo} and holography~\cite{Morgante:2022zvc} indicate that the inverse duration of the phase transition in Hubble units is $\sim 10^4$, which would lead to the signal far beyond the current detection capabilities. However, some estimates obtained with holography and AdS/CFT methods suggest that this parameter could be smaller by many orders of magnitude~\cite{Konstandin:2010cd,Ahmadvand:2017xrw,Li:2021qer}.} It was recently proposed~\cite{Breitbach:2018ddu,Fujikura:2023lkn,Zu:2023olm} to interpret the detection of stochastic gravitational background by the NANOGrav collaboration~\cite{NANOGrav:2023gor} as a signal from twin QCD FOPT. Nevertheless, this interpretation requires inverse duration time of the transition in Hubble units $\lesssim 10 $ to fit the measured data. Such small value disagrees with most accessible estimates by many orders of magnitude and, thus, it is unlikely that there is any correlation between the QCD-like FOPT that appears as a byproduct of SNR and the NANOGrav results.

\section{Supersymmetric Twin Higgs}\label{sec:SUSYTH}

Twin Higgs model is an IR effective theory and in general requires UV completion that stabilizes the scale of global symmetry breaking $f_0$ and dynamically generates the scalar potential. In this paper we focus on SNR in SUSY UV completions of TH~\cite{Falkowski:2006qq,Chang:2006ra,Craig:2013fga,Katz:2016wtw, badziak2017S,badziak2017M,badziak2017A}. SNR in composite Twin Higgs models was analyzed in~\cite{Matsedonskyi:2020kuy}.

SUSY TH models can be effectively reduced to the two copies of MSSM with the approximate discrete $\mathbb{Z}_2$ symmetry relating couplings and soft terms in each sector. We consider UV extensions where the $\mathbb{Z}_2$ symmetry between the MSSM and Twin sectors is explicitly broken by fermion Yukawa couplings. As we show in section ~\ref{sec:LSR}, the enhancement of the twin fermion Yukawas can increase the SM EW phase transition temperature, which opens a way for interesting scenarios of baryon asymmetry production. Moreover, enhanced twin Yukawas can reduce the number of relativistic species down to the level consistent with the cosmological constraints~\cite{Barbieri:2016zxn,Barbieri:2017opf}. Finally, the $\mathbb{Z}_2$ symmetry breaking between SM and twin couplings can make scalar phase transition first-order~\cite{Badziak:2022ltm}. These possibilities are extensively analyzed further in this section.

\subsection{Model description}\label{sec:Model}

Effectively, all the scenarios discussed in this section consist of two mirror copies of the MSSM. Since our analysis does not extend far beyond the TH scale $f_0$, we neglect all the particle species with masses above this scale. The LHC data implies that in this regime, all the squarks and gauginos (except bino) must be decoupled. For simplicity, we also assume $m_A=2$ TeV, which guarantees that the MSSM-like Higgses also decouple and therefore the only remaining dynamical Higgs fields are SM and Twin Higgses,
\begin{equation}
    h_I=H^u_I \sin\beta_I + H^d_I \cos\beta_I \,,
\end{equation}
where the up-type Higgs doublet $H^u_I$ couples to up-type fermions while the down-type Higgs doublet $H^d_I$ couples to down-type fermions and $ \beta_{I}\equiv\arctan\left(\langle H^u_I\rangle/\langle H^d_I\rangle\right)$. Furthermore, we take the Higgsino mass $\mu_{h}\sim1$ TeV in order not to affect the dynamics of phase transition but light enough to neglect its impact on the naturalness of the EW scale.

Below the TH scale, the only SUSY particles that directly contribute to the thermal evolution of the SM Higgs field value are the scalar partners of charged leptons and neutrinos. We call the former charged sleptons and the latter sneutrinos. The Yukawa couplings of neutral and charged leptons in the SM sector are very weak and therefore do not impact the scalar thermodynamics. However, charged slepton and sneutrino D-terms, which scale with the weak and electromagnetic gauge couplings, at high temperatures can induce corrections to the Higgs potential. Although, these new contributions could alter the electroweak transition temperature relative to the SM prediction, they are too weak to change the order of SM EW phase transition. On the other hand, a direct $\mathbb{Z}_2$-breaking in the Yukawa sector allows us to enhance some of the twin fermion-to-Higgs couplings w.r.t. their SM values. The strong enhancement of twin fermion Yukawas creates dominant contribution to the twin Higgs dynamics and indirectly affects the restoration of the SM EW symmetry.

Large hard $\mathbb{Z}_2$-breaking in the Yukawa couplings generically leads to additional fine-tuning (FT), which we quantify in appendix~\ref{app:FT}. This problem can be resolved by introducing light SUSY partners. The LHC bounds do not allow for scalar quarks below the TeV scale in the SM sector~\cite{particle2022,ATLAS:2020syg,ATLAS:2021twp}. On the other hand, direct production of charged sleptons is still poorly constrained by the LHC in some parts of the parameter space and the strongest generic experimental bounds on charged slepton masses come from LEP and are $\mathcal{O}(100)$~GeV~\cite{CMS:2024gyw,ATLAS:2024fub,ALEPH:2001oot,DELPHI:2003uqw,OPAL:2003nhx,L3:2003fyi}. Thus, we introduce charged sleptons and their twins with small soft masses, which allows us to enhance twin lepton Yukawas without violating the naturalness. 

The tree-level diagonal entries of the charged-slepton mass matrix read\footnote{The charged-slepton and sneutrino mass terms proportional to gauge couplings stem from the D-term potential.}
\begin{align}\label{eq:m_sl}
    &\hat{m}^2_{\text{sl}\;R}=\hat{\mu}^2_{e}+\frac{1}{2}\tilde{y}_{l}^2  h_B^2\cos^2{\beta_{B}}-\frac{1}{4} g'^2 h_B^2 \cos (2\beta_B),\\
    &\hat{m}^2_{\text{sl}\;L}=\hat{\mu}^2_{L}+\frac{1}{2}\tilde{y}_{l}^2  h_B^2\cos^2{\beta_{B}}-\frac{1}{8} (g^2-\,g'^2) h_B^2 \cos (2\beta_B).
\end{align}
Here $\hat{\mu}^2_{L}$ and $\hat{\mu}^2_{e}$ are the soft SUSY breaking masses of the left-handed (LH) and right-handed (RH) charged slepton twins, while $\tilde{y}_l^2\equiv\hat{y}_l^2(\tan^2 \beta+1)$ is the supersymmetric twin lepton Yukawa. Although there could, in principle, be some non‑zero mixing between the charged sleptons, for simplicity we assume it to be negligible.

\subsubsection*{Broken gauged $\mathbf{B-L}$}\label{sec:B-L}

Let us now discuss an extension of the minimal scenario in which $B-L$ and twin $B'-L'$ symmetries are gauged. In this setup the minimal model is augmented by three right-handed neutrinos (RHNs) and their twins to realize the type-I seesaw mechanism~\cite{Minkowski:1977sc,Yanagida:1980xy,Gell-Mann:1979vob}. In this case, the SM RHNs must be heavy due to a large $B-L$-breaking Majorana mass. If the twin $B'-L'$ symmetry is unbroken, the Majorana mass terms of the TS RHNs vanish, leading to the asymmetry in the effective neutrino Yukawas. An example of a UV model where $B-L$ is broken while $B'-L'$ is not is presented in appendix~\ref{app:B-L}. This scenario is particularly well-motivated because for sufficiently large twin-neutrino Yukawa couplings the dark-radiation problem is significantly relaxed, see the discussion in section~\ref{sec:N_eff}. At the same time, twin RHNs help to obtain SNR. Moreover, twin RH sneutrinos are much lighter than SM RH sneutrinos because only the latter obtain large supersymmetric masses from the Majorana terms. Therefore, the soft squared masses of twin RH sneutrinos can be very small and even negative without affecting any experimental constraints nor naturalness. 

Physical twin sneutrino masses are
\begin{align}\label{eq:m_sn}
    &\hat{m}^2_{\text{sn}\;R}=\hat{\mu}^2_{n}+\frac{1}{2}\tilde{y}_{n}^2  h_B^2\sin^2{\beta_{B}},\\
    &\hat{m}^2_{\text{sn}\;L}=\hat{\mu}^2_{L}+\frac{1}{2}\tilde{y}_{n}^2  h_B^2\sin^2{\beta_{B}}+\frac{1}{8} (g^2+\,g'^2) h_B^2 \cos (2\beta_B),
\end{align}
with $\tilde{y}_{n}^2\equiv\hat{y}_{n}^2(\cot^2 \beta+1)$ and $\hat{\mu}^2_{n}$ being the squared twin RH sneutrino soft mass which can be even negative as long as its absolute value is smaller than the supersymmetric contribution to the RH sneutrino mass from Yukawa interactions. It is remarkable that twin sneutrinos not only cancel radiative contributions to the twin Higgs mass from twin neutrinos but also could positively contribute to the SNR, which makes this model particularly attractive.    

Generically, spontaneous breaking of the $B-L$ gauge symmetry at some high energy scale induces extra D-term contribution to MSSM soft masses~\cite{Drees:1986vd,Hagelin:1989ta,Kawamura:1993uf}. If the extra contribution is $\mathcal{O}(1)$~TeV it is also possible to have light twin RH squarks in the spectrum
\begin{align}
    &\hat{m}^2_{\text{su}\;R}=\hat{\mu}^2_{u}+\frac{1}{2}\tilde{y}^{(i)}_{u}{}^2 h_B^2\sin^2\beta  +\frac{1}{6} g'^2 h_B^2 \cos (2\beta_B)  ,\\
    &\hat{m}^2_{\text{sd}\;R}=\hat{\mu}^2_{d}+\frac{1}{2}\tilde{y}^{(i)}_{d}{}^2h_B^2\cos^2\beta  +\frac{1}{12} g'^2 h_B^2 \cos (2\beta_B),
\end{align}
where $\hat{\mu}^2_{u}$ and $\hat{\mu}^2_{d}$ are the soft mass terms for the twin up-type and down-type squark while $(\hat{y}^{(i)}_{u})^2$ and $(\hat{y}^{(i)}_{d})^2$ are the Yukawas of up and down type quarks in the $i$-th TS family. These new EW-scale sfermions can help in obtaining SNR and enhance the FOPT.

\begin{figure}[t]
    \centering

    \includegraphics[scale=1.5]{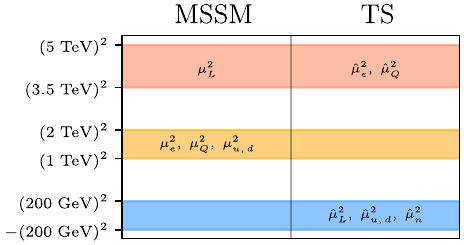}
        \caption{Schematic picture showing sfermion soft masses in the MSSM and TS for a scenario with $\mathcal{O}(1)$ TeV extra D-term contributions to the soft masses in the scenario with broken gauged $B-L$ and unbroken gauged $B'-L'$.}
    \label{fig:mass_spectrum}
\end{figure}

The extra $D$-term contribution to MSSM soft terms of sparticles charged under $B-L$ is proportional to their $B-L$ charges. Thus, light twin RH squarks impact the whole SUSY spectrum. In particular, the extra $D$-terms give large negative contribution to RH charged sleptons, so to avoid tachyons, $\mathbb{Z}_2$-symmetric soft terms for RH sleptons and their twins must be at the few TeV scale. As a consequence, twin RH charged sleptons do not contribute to SNR. Another interesting feature of SUSY spectrum is that MSSM LH and RH charged sleptons can be $\mathcal{O}(1)$~TeV so the LHC limits on them are generically satisfied in contrast to previous scenarios in which fulfilling LHC constraints requires the SUSY spectrum to have a peculiar structure. The typical pattern of sfermion masses and their twins is presented in figure~\ref{fig:mass_spectrum}.
Note that the mass splitting between the MSSM and twin stops does not affect the naturalness of the EW scale at the leading order since the $D$-term correction to the squared mass of the RH stop is opposite to that of the LH stop.

\subsection{Symmetry non-restoration in SUSY TH}\label{sec:SNRpheno}

The MSSM-like SUSY extensions introduce two types of new parameters in the mass terms: soft terms and SUSY angles ${\rm tan}\beta$. In the SM sector these parameters can take values allowed in the ordinary MSSM, and in the limit of exact mirror $\mathbb{Z}_2$ symmetry, SUSY parameters in TS are fixed. Nonetheless, the SNR requires an explicit breaking of this symmetry in the Yukawa sector, which complicates phenomenological analysis of the model.

In this paper, for simplicity we only consider $\mathbb{Z}_2$ symmetric SUSY angles 
\begin{equation}
    \tan\beta_A\simeq \tan\beta_B. 
\end{equation}
Although large renormalization group (RG) corrections from enhanced twin Yukawas favor $\tan\beta_A>\tan\beta_B$ and hence the exact mirror symmetry in SUSY angles may require some amount of tuning, an exact equality is not necessary for the TH mechanism to be at work. Moderate difference in $\tan\beta$ would only marginally affect our results by enhancing the impact of the SM top. Even though $\tan\beta$ is not the subject of direct experimental constraints, D-term SUSY UV completions that feature minimal tuning of the TH scale favor $\tan\beta\sim3$~\cite{Badziak:2017syq, Badziak:2017kjk, Badziak:2017wxn}. 

Large splitting in the Yukawa couplings, which is imposed at a scale $\Lambda \gg f_0$, radiatively induces $\mathbb{Z}_2$-breaking in the soft Higgs masses $m^2_{H_u}$ and $m^2_{H_d}$ that may have to be canceled to obtain viable SM Higgs phenomenology.
We assume that additional $\mathbb{Z}_2$-breaking in $m^2_{H_u}$ and $m^2_{H_d}$ is also generated at a scale $\Lambda$ by some unspecified UV mechanism. This cancellation may require additional FT that increases with $\Lambda$ (for quantitative results see appendix \ref{app:RGEs_res}), and therefore hereafter in this section we set $\Lambda=10^4$ GeV. We emphasize that $\Lambda$ does not have to coincide with the scale at which the MSSM and TS are UV completed. As demonstrated in appendix \ref{app:Vquarks}, the splitting in $m^2_{H_u}$, $m^2_{H_d}$, and fermion Yukawas can appear after integrating out vector-like quarks with masses $\sim\Lambda$. Beyond the vector-like quark scale, the $\mathbb{Z}_2$ symmetry is only softly broken, and the RG running of the soft Higgs parameters in the MSSM and TS is symmetric.

In the model with broken gauged $B-L$ and light twin RHNs there might be additional source of tuning stemming from the $\mathbb{Z}_2$ breaking running of the Higgs soft masses by the Yukawa couplings of RHNs. This is especially true for the standard high-scale seesaw mechanism where 
$B-L$ is broken at a scale $\mathcal{O}(10^{14})$~GeV. This problem is avoided if $B-L$ is broken at a low scale. We present a SUSY UV completion of the TH model which incorporates RHNs in appendix~\ref{app:UV_RHN}. The model is an extension of a model proposed in ref.~\cite{Badziak:2017wxn} that takes into account the gauged $B-L$ sector and allows for both low-scale and high-scale $B-L$ breaking. In our numerical analysis of FT we assume low-scale $B-L$ breaking but we comment on the scenario with high-scale $B-L$ breaking in subsection \ref{ssec:RHN}.

The remaining SM and TS SUSY couplings are assumed to be equal at $\mathbb{Z}_2$-breaking scale. The exact mirror symmetry imposed at $\Lambda$ implies that they are approximately equal also at the lower scales, except for sfermion soft mass terms whose running strongly depends on the enhanced twin Yukawas. To estimate the resultant mass splitting, we integrated full one-loop MSSM renormalization group equations (RGEs), matching the results at the scale of $\mathbb{Z}_2$ breaking. Interestingly, we found that for $\Lambda\geq 10^4$ GeV, the squares of twin slepton soft masses are generically negative at the Twin EW scale\footnote{One can obtain positive squares of the soft twin slepton masses by choosing larger slepton soft masses on the SM site.}; see appendix~\ref{app:RGEs} for details. While SUSY scalars with negative squares of the soft masses generically oppose the EW symmetry restoration at low energies (see discussion below eq. \eqref{eq:dVscal} in section~\ref{sec:SFcorr}), large negative soft terms could also dominate physical masses of the particles. Charged SUSY scalars with negative mass squared would necessarily develop non-zero VEVs that break the TS electromagnetic symmetry and generate a non-zero mass of the mirror photon. Although the non-zero mass can in principle reduce the number of the dark relativistic degrees of freedom in the model, the quantitative analysis of this scenario goes beyond the scope of this paper.

\begin{table}[t]
\centering
\begin{tabular}{|lllp{1cm}|}
\hline\hline

\multicolumn{1}{c}{\textbf{UV extension N\b{o}}} & \multicolumn{1}{c}{\hspace{2mm}$\boldsymbol{\hat{y}_u}$, $\boldsymbol{\hat{y}_c}$\hspace{2mm}} & \multicolumn{1}{c}{\hspace{2mm}$\boldsymbol{\hat{y}_s}$, $\boldsymbol{\hat{y}_b}$\hspace{4mm}} & \multicolumn{1}{c}{\textbf{EW-scale SUSY partners}} \\[1mm] \hline\hline

\multicolumn{1}{c}{\textbf{1)}}  & \multicolumn{1}{c}{$0.8$} & \multicolumn{1}{c}{$0.7$}  & \multicolumn{1}{c}{--}   \\[1mm] \hline \hline

\multicolumn{1}{c}{\textbf{2)}} & \multicolumn{1}{c}{$0.7$} &  \multicolumn{1}{c}{$0.5$}   & \multicolumn{1}{c}{ch. sleptons}\\[1mm] \hline\hline

\multicolumn{1}{c}{\textbf{3)}}  & \multicolumn{1}{c}{$0.4$} &  \multicolumn{1}{c}{$0.3$}  & \multicolumn{1}{c}{ch. sleptons, sneutrinos}   \\[1mm] \hline\hline

\multicolumn{1}{c}{\textbf{4)}}  & \multicolumn{1}{c}{$0.4$} &  \multicolumn{1}{c}{$0.3$}   & \multicolumn{1}{c}{LH ch. sleptons,}
\\
\multicolumn{1}{c}{ }  & \multicolumn{1}{c}{ } &  \multicolumn{1}{c}{ }   & \multicolumn{1}{c}{sneutrinos, RH squarks} \\[1mm]\hline\hline

\end{tabular}
\caption{Explicit $\mathbb{Z}_2$ breaking in UV extensions considered in this paper. The twin quark Yukawa couplings are set at the scale $\mu=1$ TeV. The twin down quark is assumed to be light enough to have a negligible impact on the Higgs field value evolution.
}
\label{tab:TQ}
\end{table}

Below, we consider four reference points given in table \ref{tab:TQ} with different SUSY particles that are in equilibrium with thermal plasma around the EW scale. For all of them the $\mathbb{Z}_2$ asymmetry from lepton and neutrino sectors is insufficient for the SNR and therefore some enhancement of twin quark Yukawas is necessary. Since the squark masses in MSSM and TS must be above the TeV scale, quantum correction from the quark sector to the Higgs mass parameters in the potential \eqref{eq:Vtree} cannot be fully compensated. As a result, splitting in quark Yukawas becomes a direct source of additional FT of the EW scale. In the reference points, we fix twin SM-like (hatted) Yukawa couplings, denoted by $\hat{y}_i$, so that twin quark couplings to $h_B$ do not change with $\tan\beta$. In order to make the model consistent with experimental bounds and improve its naturalness, we exploit the following ordering principles for twin quark Yukawas:
\begin{itemize}
    \item Although the naturalness favors spreading $\mathbb{Z}_2$ asymmetry across many quark species, we always keep a single twin quark light, $\sim 10 \;\text{GeV}$. While this does not worsen the tuning much, the light twin quark can be used to transfer significant part of the entropy from TS to SM and reduce the number of excessive dark relativistic degrees of freedom, as will be clarified in section \ref{sec:N_eff}.

    \item  We choose the only light twin quark to be down-type since for $\tan\beta>1$ the radiative corrections to the EW scale from down-type quark Yukawas are dominant. This is because for $\tan\beta>1$ supersymmetric Yukawas for down-type quarks are enhanced so their RG running towards large values at high energies is faster. 
    
    \item For the same reason as above, we choose somewhat smaller Yukawas for down-type quarks than those of up-type quarks. 

\end{itemize}

For each TH extension in table \ref{tab:TQ} we complete the scan in $\tilde{y}_l-\tan\beta$ plane. At every analyzed point, we use the python \texttt{CosmoTransitions} package~\cite{Wainwright:2011kj} with custom modifications to trace the minima of the effective potential \eqref{eq:V_tot} in two-dimensional field space. In the next step, we solve full one-loop RGEs in the SM and TS and find the minimal possible tuning.%
\footnote{Note that the soft masses of charged sleptons and sneutrinos in the TS are fixed in the simulations, although in principle one could fix their values in SM sector and compute the splitting at each point by solving RGEs. We found this procedure to be an unnecessary complication. In the appendix \ref{app:RGEs_res} we demonstrate that the twin slepton soft masses used by us can be easily realised for representative benchmark points and phenomenologically viable values of SM soft terms.}
The numerical results hold also beyond the EFT region with $h_A,\;h_B\ll f_0$ and the prediction for FOPT remains accurate even if the critical temperature is of the order of TH scale.

\subsubsection{Enhanced lepton Yukawas}\label{ssec:lept}

 \begin{figure}[t]
    \centering
        \subcaptionbox{\textbf{1)} from table \ref{tab:TQ}    \label{fig:SNR_ferm}}{
 \includegraphics[scale=1]{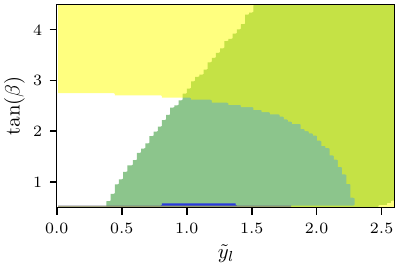}}
    \subcaptionbox{\textbf{2)} from table \ref{tab:TQ}\label{fig:SNRmu1a}}{
    \includegraphics[scale=1]{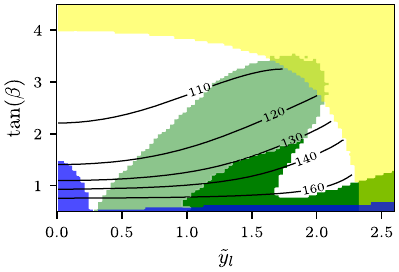}}
     \caption{The region of symmetry non-restoration (green) in $\tilde{y}_l-\tan\beta$ plane. Dark green depicts the parameter space in which SNR is followed by a FOPT in TS with $v_B(T_c)/T_c>1$. In the yellow region the Landau pole is reached below the matching scale $\Lambda=10^4\;\text{GeV}$, while in the blue region some of  squared sfermion masses are negative. \textbf{Left panel:} Scenario \textbf{1)} in table~\ref{tab:TQ} where all sfermions are assumed to have masses above $(2\;\text{TeV})^2$ and decouple. \textbf{Right panel:} Scenario \textbf{2)} in table~\ref{tab:TQ}.  Contours show the minimal possible FT, $\Delta_{v/f}$ (defined in eq.~\eqref{eq:FT_v/f}),  for a given hard $\mathbb{Z}_2$-breaking in Yukawa couplings. The TS soft masses of left-handed charged sleptons are set to $\hat{\mu}^2_{L}=(100\;\text{GeV})^2 $ while those of right-handed ones  $\hat{\mu}^2_{e}$ are varied from $-(100\;\text{GeV})^2$ to $-(250\;\text{GeV})^2$. The SM and TS squarks are heavy, $\phat{m}{}^2_{\text{sq R/L}}\gtrsim (2\;\text{TeV})^2$, and decouple.
     }
     \label{fig:SNRmu1}
\end{figure}

\noindent We begin our analysis from a basic UV extension, where all twin SUSY partners are heavy and decouple below the TH scale. The SNR is then of purely fermionic nature and requires large hard $\mathbb{Z}_2$-breaking in quark and lepton Yukawas (row \textbf{1)} in table~\ref{tab:TQ}). In figure~\ref{fig:SNR_ferm} we present the resulting scan in the $\widetilde{y}_l-\tan\beta$ plane that shows the region where the SM EW symmetry is broken a little below TH scale. As it was anticipated in section \ref{sec:FOPT}, at all the points the scalar transition was either smooth or weakly first-order, and we do not find any signs of strong FOPT. We emphasize that in this TH extension the tuning is high since there are no particles that cancel large radiative corrections from TS fermions. Therefore, it cannot be motivated by the naturalness.

If all the charged SUSY lepton partners are taken to be of order $200$ GeV rather than being heavier than the TH scale, they become a new source of $\mathbb{Z}_2$ breaking and quark Yukawas can be reduced to values given in row \textbf{2)} in table \ref{tab:TQ}. The scan in the $\widetilde{y}_l-\tan\beta$ plane for this scenario is shown in figure \ref{fig:SNRmu1a}. The contours in the plot depict the level of minimal FT of the EW scale, $\Delta_{v/f}$, stemming from the $\mathbb{Z}_2$ breaking in the Yukawa couplings (see eqs.~\eqref{eq:FT_v/f} and \eqref{eq:FT}). At each point, we adjusted soft masses of charged sleptons to maximize the SNR region and at the same time keep all the physical masses positive. The squark masses are fixed at 2 TeV, above direct experimental bounds from the LHC and not too far from $f_0$, in order to avoid large UV tuning of the TH scale. Clearly, the SNR appears in the major part of the parameter space and is often followed by strong FOPT in the TS in the region with $\tilde{y}_l>1$, $\tan(\beta)<2$. The overall model naturalness is significantly improved compared to the case with all sfermions decoupled, but still the minimal FT is worse than $1$\%. A gap in the SNR region for higher Yukawa and lower $\tan(\beta)$ values appears because, for a large effective twin slepton-to-Higgs coupling $\tilde{y}_l/\sqrt{1+\tan^2\beta}$, charged sleptons' contribution to the SM Higgs mass changes sign and opposes EW symmetry breaking.

\subsubsection{Right-handed neutrinos}\label{ssec:RHN}

 \begin{figure}[t]
    \centering
    \subcaptionbox{
    \textbf{3)} from table \ref{tab:TQ} in $\widetilde{y}_{l}-\tan\beta$ plane.
    \label{fig:SNRmu4a}}{
    \includegraphics[scale=1]{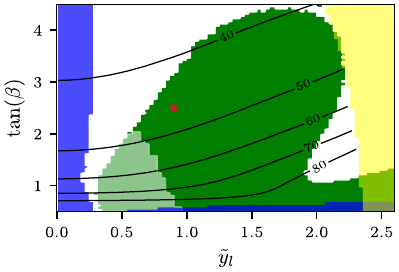}}\hspace{5mm}
    \subcaptionbox{
    \textbf{3)} from table \ref{tab:TQ} in $\widetilde{y}_{l}-\widetilde{y}_{n}$ plane 
    \label{fig:SNRmu4b}}{
    \includegraphics[scale=1]{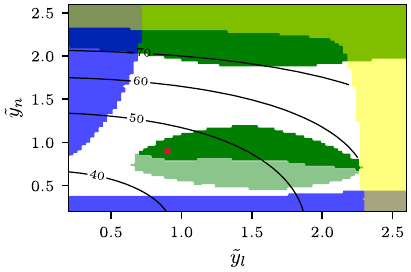}}\hspace{5mm}
        \caption{Scan across the slices of the TH parameter space with light twin RHNs and their SUSY partners for scenario \textbf{3)} in table~\ref{tab:TQ}. The color coding and contours of minimal FT are the same as in figure~\ref{fig:SNRmu1a}. On the left panel we fix $\tilde{y}_n=0.9$, while on the right panel $\tan\beta=2.5$. Soft masses of twin sleptons $\hat{\mu}^2_{L}$, $\hat{\mu}^2_{e}$, $\hat{\mu}^2_{n}$ are varied from $-(240\;\text{GeV})^2$ to $-(100\;\text{GeV})^2$. The SM and TS squarks have masses $\phat{m}{}^2_{\text{sq R/L}}\gtrsim (2\;\text{TeV})^2$ and do not affect the phase transition dynamics. The red cross marks the benchmark point for which the background field evolution is shown in figure~\ref{fig:Benchmark}.
        }
    \label{fig:SNRmu4}
\end{figure}

\noindent In section \ref{sec:B-L} we develop an extension of the model where $B-L$ and $B'-L'$ are gauged but $B-L$ is broken at some high scale while $B'-L'$ is not. TS neutrino Yukawas may be $\mathcal{O}(1)$ and the right-handed sneutrinos can have small soft mass terms. In this set-up, a substantial part of the $\mathbb{Z}_2$ breaking necessary for the SNR may be accommodated in the neutrino sector and the asymmetry in quark Yukawas can be partially relaxed, resulting in the overall reduction of FT. Taking the rest of the SUSY spectrum the same as for the scenario with light sleptons, we computed the region of SNR in the $\widetilde{y}_l-\tan\beta$ plane. The results are shown in figure \ref{fig:SNRmu4a}, where we also depict the minimal FT contours. For simplicity, we set all twin neutrino Yukawas equal. On each point in the plot we picked the soft masses of twin charged sleptons and sneutrinos to maximize the SNR region but avoid ``tachyonic'' regions with negative physical twin slepton mass.

To analyze how the SNR depends on the neutrino Yukawas we completed the complementary scan in $\widetilde{y}_{l}-\widetilde{y}_{n}$ plane (see figure~\ref{fig:SNRmu4b}). Setting the neutrino Yukawas too high causes the Landau pole to appear below $\Lambda$ scale, which is ruled out. On the other hand, if $\widetilde{y}_{n}$ are set too low the fermion-Higgs couplings in TS are too weak to prevent the EW symmetry restoration for temperatures beyond the EW scale.

In the scenario analyzed above, the SNR combines with the FOPT in a noticeably larger region than in models without RHNs, and in particular for moderate Yukawa couplings and $\tan\beta \sim 3$. Moreover, the minimal tuning can be better than $2$\%, representing a threefold improvement compared to the scenario with light charged sleptons only.

\begin{figure}[t]
    \centering
    \includegraphics[scale=0.99]{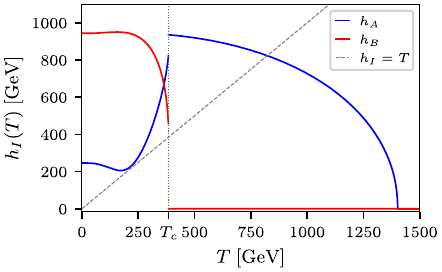}\hspace{5mm}
    \includegraphics[scale=0.99]{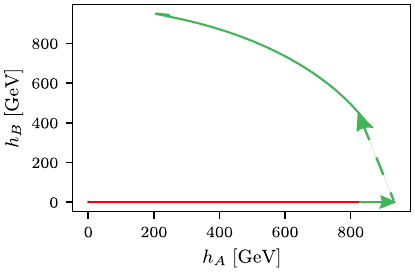}

        \caption{Left plot: VEVs during thermal evolution. Right plot: path in the $ h_{A}-h_{B}$ space with the part of curve where SM sphalerons are suppressed (i.e. $h_{A}(T)/T>1$) marked in green. The plots were completed at the test point $\tilde{y}_l=\tilde{y}_n=0.9$, $\tan\beta=2.5$, and $\hat{\mu}^2_{\nu}=\hat{\mu}^2_{\text{sl}}=-160^2$~GeV$^2$, marked by red cross in figure~\ref{fig:SNRmu4}. Baryon asymmetry wash-out does not occur below $T\approx825\; \text{GeV}$. 
        }
    \label{fig:Benchmark}
\end{figure}

For illustrative purposes, in figure \ref{fig:Benchmark} we show how the scalar VEVs change with temperature for a representative benchmark point with $\Delta_{v/f}$ below $50$ and strong FOPT. Although the plot is for a particular UV scenario, qualitatively similar picture could be obtained for any other benchmark featuring SNR and FOPT at any TH extension considered in this paper. The path in $h_A-h_B$ plane agrees well with our approximate analytical model in section \ref{sec:SFcorr}. At temperatures much above the TH scale $f_0$ the radial mode receives large thermal corrections, restoring the original $SU(4)$ of the potential. For temperatures not far above $f_0$, the SM-like Higgs field starts to develop non-zero value. Eventually, at temperatures below the $f_0$ scale, 
the evolution is bounded to the circular path, fixed by eq. \eqref{eq:EFT}. The scalar tunneling occurs in the transverse direction, where the potential barrier is elevated by the bosonic corrections to the effective potential discussed in section \ref{sec:FOPT}. 

In the above analysis we assumed a low-scale seesaw in which the $\mathbb{Z}_2$ breaking in the Yukawa couplings of RHNs is present only for scales up to $10^4$~GeV. In the high-scale seesaw the Yukawa couplings of RHNs are generated at a scale $\mathcal{O}(10^{14})$ GeV which leads to two potential problems. First, the model must be perturbative up to the seesaw scale and avoiding the Landau pole for Yukawas of RHNs below $10^{14}$~GeV sets an upper bound on $\widetilde{y}_{n}\lesssim0.5$. Such small Yukawa couplings of twin RHNs barely help in achieving SNR, as can be seen from figure~\ref{fig:SNRmu4b}.
Second, $\mathbb{Z}_2$ breaking in the Yukawa couplings of RHNs induce splitting between the SM and twin top Yukawa couplings due to long RG running up to $10^{14}$~GeV, putting the TH mechanism in danger. 

Keeping perturbativity up to $10^{14}$~GeV requires also a specific UV completion of the TH model since most models suffer from a low Landau pole scale for a coupling that gives the $SU(4)$ invariant quartic coupling. The perturbativity up to high scales can be kept only in SUSY $D$-term TH models in which the $SU(4)$ invariant quartic coupling is generated by a $D$-term of a new non-abelian gauge symmetry~\cite{Badziak:2017kjk,Badziak:2017wxn}. In Appendix~\ref{app:UV_RHN} we present a generalization of the model proposed in ref.~\cite{Badziak:2017wxn} in which RHNs with $\mathcal{O}(1)$ Yukawa couplings are charged under the new non-abelian gauge symmetry. The fact that RHNs are charged under this new gauge symmetry has several consequences that ameliorate the two problems mentioned above. First, the contribution from the extra gauge coupling $g_X$ to the beta function of the RHNs Yukawa couplings slows down its running and we found that there is no Landau pole below $10^{14}$~GeV for $\widetilde{y}_{n}$ as large as $0.9$.
Second, the Yukawa couplings of tops and RHNs are suppressed at high scales due to the effect of large gauge coupling of the new interaction on their RG running. This effect significantly reduces the FT from the $\mathbb{Z}_2$ breaking in the top sector. On the other hand, due to the RHNs being charged under the new interaction, this interaction is no longer asymptotically free and the RG running of the new gauge coupling is different in the visible and twin sector below the see-saw scale which, in turn, results in the splitting in the top Yukawa couplings. In consequence, there is some additional tuning stemming from the $\mathbb{Z}_2$ breaking in the top sector. We found that for the seesaw scale of $10^{14}$~GeV this increases $\Delta_{v/f}$ shown in figure~\ref{fig:SNRmu4} roughly by about 70.

To summarise, SNR with FOPT is possible in both low-scale and high-scale seesaw scenario. However, the high-scale seesaw requires more tuning of the EW scale. After embedding these scenarios to UV complete SUSY $D$-term TH models~\cite{Badziak:2017syq,Badziak:2017kjk,Badziak:2017wxn} we found that overall tuning of the EW scale can be as low as about $2$\% in the low-scale seesaw case. On the other hand, the high-scale seesaw requires bigger overall tuning $\mathcal{O}(0.5)$\%. The latter is not only a result of $\mathbb{Z}_2$ breaking in the top sector but also stronger upper bound from perturbativity on the gauge coupling of the new interaction  (which is no longer asymptotically free after including RHNs which are charged under it) leading to smaller $SU(4)$ invariant quartic coupling and, in consequence, to the overall tuning above the minimal one estimated by $\Delta_{v/f}$.

\subsubsection{Right-handed neutrinos and extra D-term contribution}\label{ssec:RHND}

 \begin{figure}[t]
    \centering
    \subcaptionbox{\textbf{4)} from table~\ref{tab:TQ} in $\widetilde{y}_{l}-\tan\beta$ plane. \label{fig:SNRmu3a}}{
    \includegraphics[scale=1]{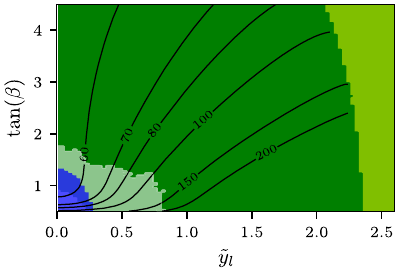}}\hspace{5mm}
    \subcaptionbox{ \textbf{4)} from table~\ref{tab:TQ} $\widetilde{y}_{l}-\widetilde{y}_{n}$ plane. \label{fig:SNRmu3b}}{
    \includegraphics[scale=1]{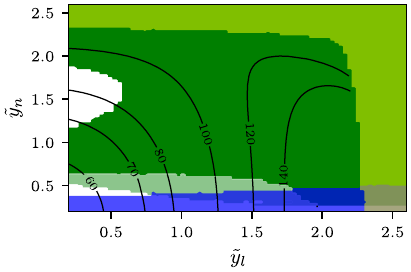}}\hspace{5mm}
        \caption{Same as figure~\ref{fig:SNRmu4} but for scenario \textbf{4)} in table~\ref{tab:TQ} with RHNs, their SUSY partners, and light twin RH squarks originating from the extra D-term contribution. On the left panel we fix $\tilde{y}_n=0.7$, while on the right panel $\tan\beta=2.5$. The results were obtained for the soft masses of slepton twins $\hat{\mu}^2_{L}$, $\hat{\mu}^2_{n}$ and RH squarks $\hat{\mu}^2_{s,\,d}$ varied form $-(240\;\text{GeV})^2$ to $-(100\;\text{GeV})^2$. The LH SM and TS squark masses are $\phat{m}{}^2_{\text{sq L}}\geq (2 \text{ TeV})^2$ and decouple.}
    \label{fig:SNRmu2}
\end{figure}

 \noindent To extend the region of SNR in the model parameter space, we also consider another scenario of $B-L$ breaking where in addition to twin RHN and their SUSY partners with $\mathcal{O}(1)$ Yukawas, twin RH squarks have $\mathcal{O}(100)$ GeV soft masses and enhance the SNR. In the MSSM extra D-terms $\Delta m^2_D$ stemming from $B-L$ breaking push RH squark masses to the few TeV scale so that the model is consistent with experimental data. Nevertheless, D-terms contribute also to the MSSM soft masses of LH squarks and charged sleptons which for TeV-scale $\Delta m^2_D$ imposes particular structure of sfermions spectrum, discussed in section \ref{sec:B-L}. When computing the minimal FT contours, we fix the appropriate sfermion spectrum by setting $\mu^2_{Q}=(2\text{ TeV})^2$, $\mu^2_{e}=(2\text{ TeV})^2$, and $\mu^2_{L}=(3.5\text{ TeV})^2$~\footnote{Low $\hat{\mu}^2_{L}$ is realized for $\mu^2_{L}\approx (3.5\text{ TeV})^2$ and therefore SM soft term was fixed to that value.} at the EW scale. The corresponding TS values are obtained with
  \begin{equation}\label{eq:splitting}
 \begin{aligned}
    &\hat{\mu}^2_{Q}=\mu^2_{Q}+\,\Delta m^2_D,\\
    &\hat{\mu}^2_{e}=\mu^2_{e}+3\,\Delta m^2_D,\\
    &\hat{\mu}^2_{L}=\mu^2_{L}-3\,\Delta m^2_D
 \end{aligned}
 \end{equation}
at the RG scale $\mu=\Lambda$, with $\Delta m^2_D$ fixed to $(2\;\text{TeV})^2$. 
Note that the $\mathbb{Z}_2$-breaking correction to the Higgs mass from $\Delta m_D^2$ is absent at the leading order, since the right- and left-handed squarks receive the soft masses with the same magnitude and opposite sign. 

The scan in figure \ref{fig:SNRmu2} shows the region of SNR in the $\widetilde{y}_l-\tan\beta$ and $\widetilde{y}_l-\widetilde{y}_n$ planes. As expected, the region of SNR is notably bigger than in the case with TeV-scale RH twin squarks and the SNR is realised for $\mathcal{O}(0.5)$ twin neutrino and twin quark Yukawa couplings. Interestingly, large twin lepton Yukawa couplings are no longer necessary to achieve SNR and in fact their small values are preferred by naturalness because large values induce FT originating from heavy twin RH sleptons. The overall enhancement of parameter space with SNR appears as there are more particles coupled to $H_B$ field which are in thermal equilibrium at the EW scale. Moreover, the tuning in major part of the parameter space is improved with respect to scenarios with no light RHN, and it is only slightly worse than in the scenario without extra $D$-term contributions as long as twin lepton Yukawa coupling is small.
On the other hand, the model does not require $\mathcal{O}(100\; \text{GeV})$ charged sleptons masses in the MSSM sector, which renders the MSSM spectrum less constrained.

\subsubsection{Temperature of the EW sphalerons freeze-out }\label{sec:T_SNR}

 \begin{figure}[t]
    \centering

    \hspace{-5mm}
    \subcaptionbox{$T_{\text{SNR}}$ in scenarios \textbf{1)}$-$\textbf{4)} with $\lambda=1$.\label{fig:fScan_Sc}}
    {\includegraphics[scale=0.99]{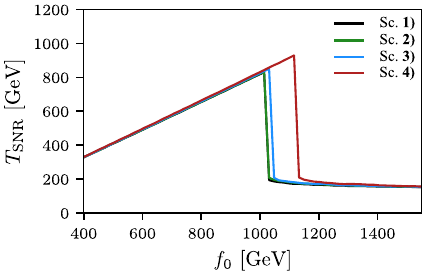}}\hspace{5mm}
    \subcaptionbox{{$T_{\text{SNR}}$ in scenarios \textbf{3)} and \textbf{4)}. Solid, dashed, and dotted lines correspond to $\lambda=1,\; 2$, and $4$, respectively.\label{fig:fScan_lam}}}
        {\includegraphics[scale=0.99]{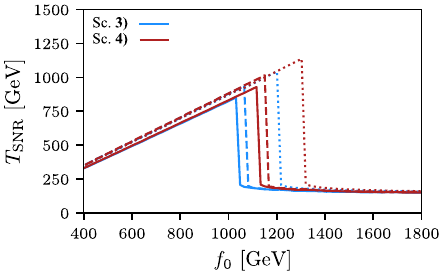}}
        \caption{The temperature $T_{\rm SNR}$ at which the EW sphalerons fall out of thermal equilibrium, plotted against the TH scale $f_0$, for scenarios \textbf{1)}$-$\textbf{4)} in table~\ref{tab:TQ}. Here we take $\tan\beta=2.0$ and $\tilde{y}_{l}=1$, and in scenarios with RHNs, $\tilde{y}_{n}=1$. For each scenario, the SUSY soft parameters were fixed to be the same as in subsections \ref{ssec:lept}-\ref{ssec:RHND}.}
    \label{fig:fScan}
\end{figure}

In section~\ref{sec:SFcorr} we demonstrated that for sufficient amount of $\mathbb{Z}_2$ breaking in the couplings of TS fermions and SUSY scalars (here indexed by $i$), the EW symmetry is not restored as long as
\begin{equation}\label{eq:val_reg}
    \max(\hat{m}_i)\lesssim T \ll f_0.
\end{equation}
 The upper bound on the region where our analytical argument holds is set by the mass of the radial Higgs mode that was integrated out in the EFT analysis, while the lower bound corresponds to the limit beyond which high-temperature approximation for thermal functions $J_{\text{B/F}}$ is not trustful anymore. The numerical results in section \ref{sec:SNRpheno}, valid also beyond the region \eqref{eq:val_reg}, confirm that EW sphalerons freeze out at high temperature $T_{\text{SNR}}\lesssim f_0$ and \emph{never} reenter thermal equilibrium, giving rise to the SNR effect.

Breakdown of the EFT approximation is controlled by the TH scale $f_0$, and hence one expects that also the $T_{\text{SNR}}$ critically depends on it. In order to quantify this relation, we computed the dependence of $T_{\text{SNR}}$ on $f_0$ at some representative benchmarks. The results are shown in figure \ref{fig:fScan}. Below, we summarize and discuss our findings.
\begin{itemize}
\item At $T\gg f_0$ both the SM and TS EW symmetries are restored. This effect is caused by the positive thermal corrections to the radial Higgs mode $\sqrt{h_A^2+h_B^2}$, which were integrated out in the EFT analysis. 

\item For fixed coupling values, the temperature of EW sphalerons freeze-out grows linearly with TH scale, with the slope common for all the considered UV extensions. 

\item As the TH scale increases, the twin fermions and their SUSY partners become heavier and decouple at higher temperatures. Early decoupling of the particles that drive the SNR leads to a gap, where neither zero temperature nor TH contributions can prevent the EW symmetry restoration, and sphaleron processes temporary reenter thermal equilibrium. Thus, linear growth of $T_{\text{SNR}}$ with $f_0$ ends at some critical TH scale value $f_0^*$, beyond which EW sphaleron processes reactivate and spoil the SNR.

\item The critical TH scale $f_0^*$, is controlled by the masses of twin particles with asymmetric couplings. The lighter are these particles at the EW scale, the stronger is their effect on the SNR at low temperatures. To raise the maximal temperature of SNR, $T_{\text{SNR}}({f_0^*})$, one needs more TS particles with weaker couplings rather than few strongly coupled SM twins, c.f. figure \ref{fig:fScan_Sc}. Alternatively, one can enhance coupling $\lambda$ in the tree-level potential \eqref{eq:Vtree}, reducing positive self-contributions to the SM Higgs mass (see figure \ref{fig:fScan_lam}).
\end{itemize}

\subsection{Relativistic degrees of freedom}\label{sec:N_eff}

At high temperatures, interactions between Higgs and its twin maintain thermal equilibrium between the visible and twin sectors. As the Universe cools down, the Higgs portal becomes inefficient and eventually the two sectors decouple. Generically, at the decoupling temperature $T_D= \mathcal{O}(1)$ GeV, there are twin degrees of freedom that are relativistic and in equilibrium with the thermal bath. After the freeze-out of the Higgs portal, a considerable amount of energy and entropy is trapped in the TS radiation. Conventionally, the energy of dark relativistic species is expressed as the excess of effective neutrino number
 \begin{equation}
     \Delta N_{\text{eff}}=\frac{\rho_{\text{TS}}}{\rho_{\nu,i}}.
 \end{equation}
Here, $\rho_{\text{TS}}$ is the energy density stored in the TS while $\rho_{\nu,i}$ is the energy density of a single SM neutrino species in the absence of dark sector. In TH models, $\Delta N_{\text{eff}}$ can be expressed as
\begin{equation}
         \Delta N_{\text{eff}}\approx\frac{4}{7}\hat{g}_{*}\big(T_{\text{CMB}})\times\Big[\frac{\hat{g}_{*}(T_D)}{g_{*}(T_D)}\frac{g_{*}(T_{\text{CMB}})}{\hat{g}_{*}(T_\text{CMB})}\Big]^{\frac{4}{3}},
\end{equation}
where $g_{*}(T)$ and $\hat{g}_*(T)$ are the numbers of degrees of freedom in the SM and Twin sector at a temperature $T$ and $T_{\rm CMB}\sim 0.3$ eV. The SM prediction for $N_{\text{eff}}$ is in very good agreement with current experimental data, which sets a stringent bound for extra relativistic degrees of freedom in BSM models $\Delta N_{\text{eff}}\equiv N^{\text{BSM}}_{\text{eff}}-N^{\text{SM}}_{\text{eff}}<0.3$ ($2\,\sigma$)~\cite{Planck:2018vyg}. TH models with $\mathbb{Z}_2$ breaking only in the Higgs masses predict $\Delta N_{\text{eff}}$ of order few, which is excluded.

One possible solution to this problem is to impose a particular pattern of $\mathbb{Z}_2$ breaking in the Yukawa sector~\cite{Barbieri:2016zxn}. Benchmark scenarios introduced in section \ref{sec:SNRpheno} inherently contain some components of this pattern since the enhancement of the twin quark and charged lepton Yukawas introduced to obtain the SNR, efficiently mitigates the fermionic sector contribution to $\Delta N_{\text{eff}}$. Further reduction of $\Delta N_{\text{eff}}$ is possible in the scenario with unbroken gauged $B'-L'$ symmetry which allows for relatively large twin neutrino masses as in this case all twin neutrinos freeze out much before the decoupling of the SM and twin sector.

Nevertheless, even if all twin fermions are sufficiently heavy to fall out of the thermal equilibrium before $T_D$, there is still a contribution to dark radiation from twin photons and gluons. The former one can be easily aligned with the current experimental constraints, while the latter requires particular treatment. Generically, TS decoupling precedes twin QCD phase transition, but by adjusting the mass of the lightest quark in the twin sector (which in our case is the twin down quark), it is possible to inverse their order. Moreover, the presence of sufficiently light twin quark(s) assures fast decay of twin glueballs that form during the confinement in TS and carry sizable fraction of the total entropy.%
\footnote{This mechanism was first described in~\cite{Barbieri:2016zxn}, although the authors considered remaining four ``heavy'' quarks lighter from the ones introduced in section~\ref{sec:SNRpheno}. In spite of this, taking larger ``heavy'' quark masses has no impact on $T_D$ while the twin $T_c^{\text{(QCD)}}$ becomes larger.}
This constrains the twin down quark mass roughly to the range $[5\;\text{GeV},20\;\text{GeV}]$ that corresponds to $\hat{y}_d\in[0.007,0.03]$ for $f_0=1$ TeV. Such small value of down Yukawa would not affect our predictions for SNR or FT, so the reduction of $\Delta N_{\text{eff}}$ can be freely accommodated in our model. The expected $\Delta N_{\text{eff}}$ can be reduced down to about $0.3$ in scenario \textbf{2)}, where quark and charged leptons leave thermal equilibrium much before $T_D$ and down to $\sim 0.14$ in scenarios \textbf{3)} and \textbf{4)} where also twin neutrinos fall out of thermal equilibrium before the MSSM and Twin sectors decouple.%
\footnote{In scenarios \textbf{3)} and \textbf{4)} we restricted our analysis to the case where $B'-L'$ gauge boson is massless, yielding additional contribution to $\Delta N_{\rm eff}$. Nevertheless, in the parameter space region where $B'-L'$ is spontaneously broken by sneutrino VEV, $\Delta N_{\rm eff}$ can be reduced down to $0.08$. Further reduction is possible when charged twin sfermion spontaneously breaks $\text{U}(1)$, rendering dark photon massive. However tracing scalar VEVs in these scenarios is much more involved than in the two-field scenario considered in this work.}
While scenario \textbf{2)} can be probed in the near future by Simons Observatory~\cite{SimonsObservatory:2018koc}, the scenario with heavy twin neutrinos will be probed by CMB-S4 experiments that are expected to constrain $\Delta N_{\text{eff}}$ with the precision around $0.05$ at 95~\%~C.L.~\cite{CMB-S4:2016ple}.

\section{Electroweak symmetry non-restoration and axiogenesis}\label{sec:SNRaxio}

In this section, we discuss the implications of the symmetry non-restoration to axiogenesis~\cite{Co:2019wyp}. In particular, we argue that the minimal axiogenesis scenario works with the SNR in Twin Higgs.

In axiogenesis, a QCD axion is introduced, and it is assumed to rotate in the field space in the early universe. The rotation can be indeed initiated by the Affleck-Dine mechanism~\cite{Affleck:1984fy}. The angular momentum of the rotation, namely, the PQ charge, is transferred into baryon asymmetry through the QCD and weak sphaleron processes. The baryon asymmetry is frozen as the weak sphaleron process freezes out. The baryon number density $n_B$ normalized by the entropy density $s$ is
\begin{align}
\label{eq:axiogenesis}
    Y_B \simeq 8.5 \times10^{-11} 
    \left( \frac{c_B}{0.1} \right) 
    \left( \frac{Y_\theta}{500}  \right)
    \left(\frac{10^8~{\rm GeV}}{f_a}\right)^2 \left( \frac{T_{\rm sph}}{130~{\rm GeV}} \right)^2,
\end{align}
where $c_B$ is a model-dependent constant that is typically $\mathcal{O}(0.1)$, $Y_\theta$ is the PQ charge density normalized by the entropy density, $f_a$ is the axion decay constant, and $T_{\rm sph}$ is the temperature at which the weak sphaleron process freezes out. The formula for $c_B$ as a function of the axion couplings can be found in~\cite{Co:2020xlh}.

The kinetic energy density of the rotation is eventually transferred into the axion dark matter density via the kinetic misalignment mechanism~\cite{Co:2019jts,Co:2021rhi,Eroncel:2022vjg}. The axion dark matter density $\rho_{\rm DM,a}$ normalized by the entropy density is
\begin{align}
\label{eq:KMM}
\frac{\rho_{\rm DM,a}}{s} = c_a m_a Y_\theta 
\simeq 0.40 {\rm eV} \times c_a
\left( \frac{Y_\theta}{7} \right)
\left( \frac{10^8~{\rm GeV}}{f_a} \right),
\end{align}
where $c_a$ is a constant that is expected to be $\mathcal{O}(1)$.%
\footnote{Here we assume that extra contributions from the acoustic misalignment mechanism~\cite{Eroncel:2025qlk,Bodas:2025eca} or parametric resonance~\cite{Co:2017mop,Co:2020dya} are at the most comparable to the kinetic-misalignment contribution.}
A recent lattice computation~\cite{Fasiello:2025ptb} finds that $c_a \simeq 0.6$ for $f_a \lesssim 10^9$ GeV. 

From eqs.~\eqref{eq:axiogenesis} and \eqref{eq:KMM},
the ratio between the dark matter and baryon densities is given by
\begin{align}
\label{eq:axion_DM}
\frac{\Omega_{\rm DM,a} h^2}{\Omega_{\rm b} h^2} \simeq 360 \ \left( \frac{f_a}{10^8~{\rm GeV}} \right) \left( \frac{130~{\rm GeV}}{T_{\rm sph}} \right)^2 \left( \frac{0.1}{c_B} \right) c_a.
\end{align}
To avoid the overproduction of axion dark matter, it is required that
\begin{align}
    f_a < 2.4\times 10^6~{\rm GeV} \frac{c_B}{0.1} \frac{0.6}{c_a} \left( \frac{T_{\rm sph}}{130~{\rm GeV}} \right)^2.
\end{align}
For the standard electroweak phase transition, where $T_{\rm sph}\simeq 130$ GeV~\cite{DOnofrio:2014rug}, the upper bound on $f_a$ is inconsistent with the astrophysical lower bound on $f_a \gtrsim $ few $ \times 10^8$ GeV~\cite{Carenza:2019pxu,Buschmann:2021juv} unless the axion is astrophobic~\cite{DiLuzio:2017ogq,Bjorkeroth:2019jtx,Badziak:2023fsc,Badziak:2024szg}. 

With symmetry non-restoration, the electroweak sphaleron process freezes out at a higher temperature and the upper bound on $f_a$ is relaxed. For example, for $T_{\rm sph} = 800$ GeV, minimal axiogenesis requires $f_a < 9\times 10^7$ GeV $\times (c_B/0.1)(0.6/c_a)$ and the astrophysical lower bound is satisfied by $c_B$ somewhat larger than $0.1$ or a mild suppression of the axion-nucleon couplings.

\section{Conclusions}\label{sec:SUM}

We have investigated high-temperature EW symmetry non-restoration in TH models. By analyzing two-field thermal evolution, we have shown that the EW symmetry can be unbroken for temperatures up to the twin EW scale, which in natural realizations of the TH mechanism is around 1~TeV. A crucial feature of the scenario are large Yukawa couplings of twin partners of the SM fermions, which effectively induce a negative thermal mass squared of the SM Higgs. 

The EW SNR requires $\mathcal{O}(1)$ twin Yukawa couplings, which correspond to large explicit $\mathbb{Z}_2$ breaking which reintroduces tuning in the Higgs scalar potential. To ameliorate this problem, we embedded the TH models in SUSY UV completion. In the visible sector, LHC constraints impose stringent lower bounds on colored sfermion masses, which in turn set limits on the masses of their twins. On the other hand, constraints for charged sleptons are significantly weaker allowing for sleptons as light as a few hundred GeV. By solving twin MSSM RGEs we showed that large twin Yukawa couplings of leptons generically induce negative soft masses squared for twin sleptons. The light twin sleptons cancel quadratically divergent contributions to the Higgs mass parameters from explicit $\mathbb{Z}_2$ breaking in the lepton Yukawa sector, thereby partially mitigating FT. Moreover, they help to achieve EW SNR and reduce the required size of twin quark Yukawa couplings, which further improves the naturalness of the model. On top of that, presence of sleptons leads to first-order phase transition across a large region in the parameter space, although the associated gravitational-wave signal is weak. This is because twin fermions that are always needed to achieve SNR counteract to the sfermion effect on creating a potential barrier between the metastable and true vacua.

The model we consider can be further extended with RHNs and their twins, assuming that $B-L$ and $B'-L'$ symmetries are gauged. The RHNs in the SM are very heavy due to large $B-L$ breaking Majorana masses, while in TS Majorana mass terms can be forbidden by unbroken $B'-L'$ symmetry. In this scenario, twin RH sneutrinos can also contribute to the SNR due to their small or negative soft masses squared. As a result, EW SNR is obtained for twin neutrino and twin lepton Yukawa couplings below one and the twin Yukawa couplings for four heavy quarks not higher than 0.4. While such values of twin quark Yukawa couplings still contribute significantly to the tuning of the EW scale, the overall FT can be lower than in the TH scenario where  charged sleptons are the only light sfermions  and in many ordinary SUSY extensions of the SM. Furthermore, in this setup, twin neutrinos with $\mathcal{O}(1)$ Yukawa couplings are sufficiently heavy so that the amount of entropy stored in the twin degrees of freedom at low temperature is significantly reduced. In particular, the effective number of neutrinos, $\Delta N_{\rm eff}$, can be brought below current experimental limits from Planck provided that the lightest twin quark mass lies roughly between 5 GeV and 20 GeV while the remaining twin quarks are significantly heavier.

We also studied the above scenario involving RHNs under the additional assumption of a sizable $D$-term contribution to the MSSM soft masses, arising from the spontaneous breaking of gauged $B-L$. These new $D$-terms induce a splitting in the MSSM sfermion soft masses with different chiralities. As a result, the RH twin squarks in TS can be also very light while the MSSM RH squarks remain heavy, as only the latter receive large $D$-term contributions, see~figure~\ref{fig:mass_spectrum} for a schematic spectrum. The presence of light twin RH squarks enables EW SNR and FOPT even without enhancing twin lepton Yukawa couplings and twin neutrino Yukawas above $0.5$. The additional tuning of the EW scale in this model variant is only slightly worse than in the case without extra $D$-term contributions, provided that twin lepton Yukawa couplings are small. A key advantage of this scenario is that electroweak SNR and FOPT can be realized while keeping all MSSM sparticles above the TeV scale, and the LHC constraints are easily satisfied without any further assumptions.

The increased temperature at which the EW symmetry breaks down has important implications for minimal axiogenesis which may explain the baryon asymmetry and DM abundance by axion rotation in the field space. This is because earlier decoupling of EW sphaleron processes leads to a larger axion decay constant for which DM and baryon abundances fit the observed values. We showed that temperatures of EW sphaleron decoupling that can be reached without excessive tuning of the EW scale enable the successful minimal axiogenesis with the axion decay constant above $10^8$~GeV, relaxing the tension of this scenario with astrophysical constraints on axion couplings. As a result, the DM puzzle, unknown origin of the baryon asymmetry together with the EW hierarchy and strong CP problems are all resolved in a relatively simple extension of the SM. 

EW symmetry non-restoration can also impact other baryogenesis mechanisms. For example, we may consider a scenario where a first-order twin EW phase transition produces a twin baryon asymmetry that is converted into a SM baryon asymmetry via some portal interactions. Depending on the nature of the portal interaction, EW symmetry non-restoration may be necessary. We leave the investigation of other baryogenesis scenarios for a future work.

\section*{Acknowledgments}
We would like to thank Raymond Co, Oleksii Matsedonskyi, Bogumiła Świeżewska and Isaac Wang for useful discussions and correspondence. The work of MB and IN was partially supported by the National Science Centre, Poland, under research grant no. 2020/38/E/ST2/00243. The work of KH was partially supported by a DOE grant DE-SC0025242; a Grant-in-Aid for Scientific Research from the Ministry of Education, Culture, Sports, Science, and Technology (MEXT), Japan 20H01895; and World Premier International Research Center Initiative (WPI), MEXT, Japan, Kavli IPMU.

\appendix

\numberwithin{equation}{section}

\section{Leading order corrections to the effective potential}\label{app:loop}
\subsection{One-loop corrections at zero and finite temperature}\label{app:1loop}

At zero temperature, one-loop contributions to \eqref{eq:Vtree} are given by~\cite{Coleman:1973jx}
\begin{equation}
 V_{\text{CW}}(h_A,\;h_B)=\sum_{i\in\text{particles}}\int \frac{d^4p}{2\pi^4}\log\big[p^2+m^2_i(h_A,\;h_B)\big],
\end{equation}
where the integral is taken in the Euclidean momentum space and the sum runs over the particle content of the TH model.
After regularization and renormalization, we obtain
\begin{equation}\label{eq:VCW}
\begin{aligned}
V_{\text{CW}}(h_A,\;h_B)=\sum_{i\in\text{bosons}}\frac{n_{i} }{64 \pi^2} m^4_{i}(h_A,\;h_B)\left(\log (\frac{m^2_{i}(h_A,\;h_B)}{\mu^2})-c_i\right)\\
-\sum_{i\in\text{fermions}} \frac{n_{i} }{64 \pi^2}m^4_{i}(h_A,\;h_B)\left(\log \frac{m^2_{i}(h_A,\;h_B)}{\mu^2}-\frac{3}{2}\right),
\end{aligned}
\end{equation}
where $n$ is the number of the degrees of freedom of the specie $i$ and the coefficients $c_i$ are equal to $5/6$ for vectors and $3/2$ for scalars. In the Landau gauge, Nambu-Goldstone modes also run in the loops and therefore one should formally account for them in the sum. However, loop expansions mix Nambu-Goldstone contributions from different orders. After the resummation of their leading-order contribution~\cite{Martin:2014bca, Elias-Miro:2014pca} they have no impact on the position or depth of the true vacuum, and therefore can be dropped. Throughout this paper we use the MSbar renormalization scheme with a renormalization scale $\mu=f_0$. We stress that the one-loop effective potential \eqref{eq:VCW} and all the observables derived from it depend on the choice of the renormalization scale $\mu$. This is unfortunately unavoidable if one works with the leading order approximation of the effective potential. However, in~\cite{Badziak:2022ltm} we show that in TH the results are generally stable under shift of the renormalization scale.\footnote{The stability under the change of the renormalization scale is checked in the UV agnostic version of TH, with $\mathbb{Z}_2$ breaking only in the scalar potential. The $\mu$ is shifted from $200\text{ GeV}$ to $\,2000\text{ GeV}$, and the corresponding change in the critical temperature is below $5\%$.}

At finite temperature, the one-loop corrections to \eqref{eq:Vtree} are obtained by evaluating the Coleman-Weinberg integral in the imaginary-time formalism~\cite{Dolan:1973qd}. The result splits into zero and finite temperature pieces, with all the infinities captured by the zero temperature part, equivalent to \eqref{eq:VCW}. The temperature-dependent part is finite and reads
\begin{equation}\label{eq:V_therm}
V_{\text{therm}}(h_A,\;h_B)=\frac{T^4}{2\pi^2}\sum_{i\in\text{bosons}}n_i\text{J}_{B}(\frac{m^2_{i}}{T^2})-\frac{T^4}{2\pi^2}\sum_{i\in\text{fermions}}n_i\text{J}_{F}(\frac{m^2_{i}}{T^2})
\end{equation}
with
\begin{equation}\label{eq:JBJF}
    \begin{aligned}
J_B(x)=\int_0^{\infty}dk\, k^2 \log[1-\exp(-\sqrt{k^2+x})],\\
J_F(x)=\int_0^{\infty}dk\, k^2 \log[1+\exp(-\sqrt{k^2+x})].
\end{aligned}
\end{equation}
In the Landau gauge, one must account for Nambu-Goldstone modes in the thermal part of the effective potential. The tree-level Nambu-Goldstones masses squared are negative whenever the tree-level potential is concave, giving rise to the imaginary contributions to $V_{\rm tot}$. While imaginary terms usually signal the instability of the effective potential, there is no instability here~\cite{Athron:2023xlk}. We remedy this problem by only including the leading order terms in high temperature expansion of the Nambu-Goldstone thermal functions, which does not lead to imaginary contributions. This approach is motivated by explicit computation in the Abelian Higgs model that compares observables derived from the effective potentials computed in different gauges~\cite{Arnold:1991uh}, and shows that the leading order term in $m^2/T^2$ is gauge invariant.

\subsection{Resummation of infrared divergences }\label{app:resummation}

Finite-temperature loop expansion of the effective potential is not always equivalent to the expansion in coupling constants. This is because certain classes of infrared-divergent diagrams yield contributions to the effective potential that do not get suppressed by the loop factor~\cite{Linde:1980ts}. To complete the consistent expansion in couplings, one needs to reorganize the perturbative series, resummating the most relevant infrared-divergent diagrams. At the leading order, this can be done by replacing the field dependent masses in the potential \eqref{eq:V_tot} with their thermal analogues~\cite{Parwani:1991gq, Arnold:1992rz}
\begin{equation}
    m^2_{i}(h_I)\rightarrow\overline{m}_{i}^2\equiv m^2_i(h_I)+\Pi(T).
\end{equation}
Here $\Pi(T)$ is the thermal contribution that for the SUSY models can be obtained from more general expressions, given in~\cite{Comelli:1996vm}.
If the particles feature kinetic mixing, the thermal corrections are added to the diagonal terms of the mass matrix written in the interaction basis. For instance, the resummed thermal masses of the SM and TS Higgs fields are the eigenvalues of
\begin{equation}
       \overline{M}_{h,\;H}=\begin{pmatrix}
3(\lambda+\kappa)h_A^2+\lambda h_B^2-f_0^2(\lambda-\sigma) +\Pi_{h_A} & 2\lambda h_A h_B \\
2\lambda h_A h_B & 3(\lambda+\kappa) h_B^2+\lambda h_A^2-f_0^2\lambda +\Pi_{h_B}
\end{pmatrix}.
\end{equation}

Below, we list the expressions for $\Pi(T)$ used in the numerical computations. As in the main text, we use here hat to denote TS particles and couplings. The hats are omitted for the twin Higgs, $g$, and $g'$.

 \begin{itemize}
    \item \textbf{Higgs sector}\\
The thermal corrections to the masses of the SM Higgs and its twin read
\begin{equation}
\begin{aligned}
&\Pi_{h_A}=\Pi^{(\text{TH})},\\
&\Pi_{h_B}=\hat{\Pi}^{(\text{TH})}+\hat{\Pi}^{(\bcancel{\mathbb{Z}_2})}.
\end{aligned}
\end{equation}
Here, $\Pi^{(\text{TH})}$ is the ordinary mirror TH thermal contributions to Higgs mass in the Landau Gauge~\cite{Fujikura:2018duw}
\begin{equation}
\phat{\Pi}^{(\text{TH})}=\Big(\frac{5}{6}\lambda+\frac{1}{2}\kappa+\frac{1}{16}g'^2+\frac{3}{16}g^2+\frac{1}{4} \phat{y}_t{}^2\Big)\,T^2,
\end{equation}
which includes the contributions from the SM and TS Nambu-Goldstone modes.
The $\hat{\Pi}^{(\bcancel{\mathbb{Z}_2})}_{h_B}$ appears due to hard mirror $\mathbb{Z}_2$-breaking in the twin sector
\begin{equation}
\begin{aligned}
    \hat{\Pi}^{(\bcancel{\mathbb{Z}_2})}=\frac{1}{24}\bigg[&\frac{1}{2}\sum_{\text{Fermions}}\big(\hat{n}_F \hat{y}_F^2-n_F y_F^2 \big)+\big(\hat{n}_{\rm su}\,\hat{y}^2_u-n_{\rm su}\,y^2_u\big)+\big(\hat{n}_{\rm sd}\,\hat{y}^2_d-n_{\rm sd}\,y^2_d\big)\\
    &+\big(\hat{n}_{\text{sl}}\,\tilde{y}^2_l-n_{\text{sl}}\,y^2_l\big)\cos^2\beta +\hat{n}_{\text{sn}}\,\tilde{y}^2_n \sin^2\beta\bigg]T^2.
\end{aligned}
\end{equation}

The sum in the above expression runs over all fermions, whose twin Yukawa couplings were enhanced with respect to their SM values. The coefficients $\phat{n}_F$, $\phat{n}_{\text{su}}$, $\phat{n}_{\text{sd}}$, $\phat{n}_{\text{sl}}$, and $\hat{n}_{\text{sn}}$ denote, respectively, the total numbers of fermion, up-type RH squark, down-type RH squark, charged slepton, and sneutrino degrees of freedom that are in equilibrium with the thermal plasma.
    \item \textbf{Gauge sector}\\
The longitudinal modes of gauge vectors also receive thermal corrections
\begin{equation}
    \phat{\Pi}_W=\phat{\Pi}_Z=\frac{11}{6}g{\,}^2T^2+\phat{\Pi}^{\text{(SUSY)}}_W,\quad\qquad\phat{\Pi}_{\gamma}=\frac{11}{6}g'{}^2T^2+\phat{\Pi}^{\text{(SUSY)}}_\gamma.
\end{equation}
The second terms are given by
          \begin{equation}
        \phat{\Pi}^{\text{(SUSY)}}_W=\frac{1}{24}\phat{n}_{\text{sl}}\,g^2T^2,\quad\phat{\Pi}^{\text{(SUSY)}}_\gamma=\Big(\frac{1}{8}\phat{n}_{\text{sl}}+\frac{2}{27}\hat{n}_{\rm su}+\frac{1}{54}\hat{n}_{\rm sd}\Big)g'{}^2T^2.
     \end{equation}
Transverse modes of vector gauge bosons do not receive thermal corrections at the leading order.
     \item \textbf{SUSY scalars}\\
The thermal masses of charged sleptons, sneutrinos, RH sneutrinos, and RH squarks are 
\begin{equation}
\begin{aligned}
\hat{\Pi}_{\text{sl}\;R}=&\Big\{\frac{1}{72}g'^2\big[24-(2\hat{n}_{\text{su}}-\hat{n}_{\text{sd}})-6\cos(2\beta)\big]+\frac{1}{6}\tilde{y}_{l}^2(1+\cos^2\beta)\Big\}T^2,\\[2mm]
\hat{\Pi}_{\text{sl}\;L}=&\Big\{\frac{1}{4} g^2+\frac{1}{144} g'^2\big[12+(2\hat{n}_{\rm su}-\hat{n}_{\rm sd})+6\cos(2\beta)\big]+\frac{1}{12}\tilde{y}_{l}^2(1+\cos^2\beta)\\
&+\frac{1}{12}\tilde{y}_{n}^2(\sin^2\beta+\theta_{n})\Big\}T^2,\\[2mm]
\hat{\Pi}_{\text{sn}\;R}=&\frac{1}{6}\tilde{y}_{n}^2(1+\sin^2\beta)T^2,\\[2mm]
\hat{\Pi}_{\text{sn}\;L}=&\Big\{\frac{1}{4} g^2+\frac{1}{144} g'^2\big[12+(2\hat{n}_{\rm su}-\hat{n}_{\rm sd})+6\cos(2\beta)\big]+\frac{1}{12}\tilde{y}_{l}^2(1+\cos^2\beta)\\
&+\frac{1}{12}\tilde{y}_{n}^2(\sin^2\beta+\theta_{n})\Big\}T^2,\\[2mm]
\hat{\Pi}_{\text{su}\;R}=&\Big\{\frac{4}{9}g_s^2+\frac{1}{108} g'^2\big[16 +(2\hat{n}_{\rm su}-\hat{n}_{\rm sd})+6\cos(2\beta) \big]+\frac{1}{6}\hat{y}_u^2 \Big\}T^2,\\[2mm]
\hat{\Pi}_{\text{sd}\;R}=&\Big\{\frac{4}{9}g_s^2+\frac{1}{216} g'^2\big[8 -(2\hat{n}_{\text{su}}-\hat{n}_{\text{sd}})-6\cos(2\beta) \big]+\frac{1}{6}\hat{y}_d^2 \Big\}T^2.
\end{aligned}
\end{equation}
Here, $\theta_{n}$ is unity if RHNs from a given generation are equilibrium with thermal plasma and zero if they are decoupled.
 \end{itemize}

\section{Fine-tuning from $\mathbb{Z}_2$-breaking Yukawas}\label{app:FT}

In the TH models where PNGB protection of the Higgs mass is combined with a mechanism that protects the global symmetry-breaking scale $f_0$, the FT of EW scale can be split into two components: the tuning of the SM Higgs VEV $v$ against $f_0$, $\Delta_{v/f}$, and the tuning of $f_0$ in the UV-completion, $\Delta_{f}$. As long as $\Delta_{v/f}$ is independent of $\Delta_{f}$ the total tuning is a product of these two components,
\begin{equation}
    \Delta_v=\Delta_{v/f}\times\Delta_{f}.
\end{equation}
 While in general both factors can be relevant, it is mostly $\Delta_{v/f}$ that is sensitive to mirror symmetry breaking. In the mirror TH with minimal breaking of mirror symmetry, the IR part of the tuning can be computed analytically~\cite{Craig:2013fga},
\begin{equation}\label{eq:FT_min}
    \Delta^{(\text{min})}_{v/f}\approx\frac{f^2_0}{2v^2}-1.
\end{equation}
An explicit mirror symmetry breaking in the Yukawa sector generically contributes to the $\mathbb{Z}_2$-breaking term $f_0^2\sigma$ in the scalar potential. When twin Yukawas are much larger than their SM analogues, the $\sigma$ value is overshot, and the contribution from twin fermions has to be canceled by a tree-level term. In such a case, the IR part of the tuning is larger than that in \eqref{eq:FT_min}, and for SUSY TH it can be estimated by%
\footnote{This expression can be derived from the eq.~(6) in~\cite{Barbieri:2016zxn} by a direct computation of the logarithmic derivative with respect to $\Delta m^2$.}
\begin{equation}\label{eq:FT_v/f}
    \Delta_{v/f}\approx\frac{\Delta m^2}{\frac12 m_h^2},
\end{equation}
with 
\begin{equation}\label{eq:Delta_m2}
    \Delta m^2\equiv\left|\Delta m_{H_u}^2\sin^2\beta+\Delta m_{H_d}^2\cos^2\beta\right|,
\end{equation}
where $\Delta m_{H_u}^2$ ($\Delta m_{H_d}^2$) is defined as a difference between the MSSM and TS soft mass of the up-type (down-type) Higgs.%
\footnote{While it is not obvious whether one should compute $\Delta m_{H_u}^2$ and $\Delta m_{H_d}^2$ at RG scale $\mu=v_{\text{EW}}$ or $\mu=f_0$, in the considered UV extensions we found difference between these two approaches rather small, below $10\%$. For numerical computation we take SM Higgs masses at $\mu=v_\text{SM}$ while in TS Higgs masses are taken at $\mu=f_0$.}

In the most natural D-term UV completions of the TH model, $\Delta_f$ can be close to unity~\cite{Badziak:2019kuo}, and thus $\Delta_{v/f}$ represents the minimal possible tuning of a TH model with a given amount of $\mathbb{Z}_2$-breaking
\begin{equation}\label{eq:FT}
\Delta^{(\text{min})}_v \approx \Delta_{v/f}.
\end{equation}
For the TH variants considered in this paper, minimal FT can only be achieved if the enhancement of twin Yukawas does not require an unnatural UV completion compared to the scenario with minimal $\mathbb{Z}_2$-breaking. In such cases, $\Delta_f$ remains roughly the same as in the SUSY TH with minimal $\mathbb{Z}_2$-breaking. Therefore, throughout the paper we discuss representative mechanisms that could naturally generate the asymmetry between the twin and SM Yukawa couplings required for the SNR.

\section{$\mathbb{Z}_2$ breaking in the $B-L$ breaking sector}\label{app:B-L}

In this appendix, we show that the $B-L$ symmetry is broken while the twin $B'-L'$ symmetry is unbroken in a simple setup. We consider the following superpotential
\begin{align}
W = \lambda X (\phi \bar{\phi} + \phi' \bar{\phi}' - v^2),
\end{align}
where the $U(1)_{B-L}\times U(1)_{B-L}'$ charges of the chiral fields are given by $X(0,0)$, $\phi(1,0)$, $\bar{\phi}(-1,0)$, $\phi'(0,1)$, and $\bar{\phi}'(0,-1)$. The F term of $X$ fixes other chiral fields on the moduli space $\phi \bar{\phi} + \phi' \bar{\phi}' = v^2$ to break $U(1)_{B-L}\times U(1)_{B-L}'$.

Let us first discuss the case where $B-L$ and $B'-L'$ are global symmetries.
The potential has an accidental $SU(2)$ symmetry, under which $(\phi, \phi')$ and $(\bar{\phi}, \bar{\phi}')$ are doublets. Because of the $SU(2)$ symmetry, the vacuum is degenerated.
The accidental $SU(2)$ symmetry is preserved also by soft supersymmetry breaking at the tree-level,
\begin{align}
V_{\rm soft} = m^2\left(|\phi|^2 + |\phi'|^2\right) + \bar{m}^2 \left(|\bar{\phi}|^2 +  |\bar{\phi}'|^2\right).
\end{align}
However, quantum corrections can break the $SU(2)$ symmetry,
\begin{align}
V_{\rm soft} = \Delta m^2|\phi|^2 {\rm ln}\frac{|\phi^2|}{\mu^2} + \Delta m^2|\phi'|^2 {\rm ln}\frac{|\phi'|^2}{\mu^2} + \Delta \bar{m}^2|\bar{\phi|}^2 {\rm ln}\frac{|\bar{\phi}^2|}{\mu^2} + \Delta \bar{m}^2|\bar{\phi}'|^2 {\rm ln}\frac{|\bar{\phi}'|^2}{\mu^2} 
\end{align}
If $\Delta m^2 + \Delta \bar{m}^2 <0$, there are two vacua, $(\phi\bar{\phi},\phi'\bar{\phi}')= (v^2,0)$ and $(0,v^2)$, where $\mathbb{Z}_2$ is spontaneously broken. Negative $\Delta m^2$ may be generated by quantum corrections by Yukawa interaction of $\phi$s and negative scalar soft masses.

When $B-L$ and $B'-L'$ are gauged, the D-term potential breaks the $SU(2)$ symmetry, but still there is a flat direction that satisfies
\begin{equation}
    \phi = \bar{\phi},~\phi' = \bar{\phi}',~ \phi^2 + \phi'^2=v^2.
\end{equation}
The tree-level soft mass only requires that the phases of $\phi$ and $\phi'$ vanish and keeps the direction $|\phi|^2 + |\phi'|^2$ flat. The flatness is lifted by quantum corrections to the soft mass, and if $\Delta m^2 + \Delta \bar{m}^2 <0$, there are two $\mathbb{Z}_2$-breaking vacua mentioned above. 
Negative $\Delta m^2$ may be generated by the renormalization by the $B-L$ gauge interaction and the $B-L$ gaugino mass.

\section{Running of the soft SUSY-breaking parameters}\label{app:RGEs}
At low energies, the SUSY TH UV extensions proposed in this work effectively reduces to two copies of the MSSM (in some scenarios augmented with RHN sector) coupled through the scalar portal. In the UV extensions without light RHNs, the running of the couplings and masses can be approximately computed by solving the MSSM RGEs in each sector and matching the solutions at $\mathbb{Z}_2$-breaking scale $\Lambda$. The same procedure can be followed in the model including twins RHNs, but the RGEs in the twin sector must be modified to account for their presence. (This model is usually referred in the literature as MSSM+3RHN.) In this appendix, we list the one-loop RGEs that cover all of the UV extensions analyzed in this paper\footnote{For conciseness, we write explicitly the RGEs for MSSM+3RHN. The MSSM equations are recovered by setting RHN Yukawas and trilinear couplings to zero.} and solve them numerically. The solutions are used to show how the splitting between the soft slepton mass and the minimal tuning depend on the matching scale $\Lambda$.

\subsection{One-loop RGEs}\label{app:RGEs_eq}

The one-loop RGEs for the coupling constants form a system of coupled first-order non-linear differential equations. Some of these equations form closed subsystems, which can be solved independently. The solution to the simplest subsystem describing the running of gauge couplings and Gaugino masses is available e.g. in.~\cite{MSSMWorkingGroup:1998fiq}. This solution can be applied without any modifications to the SM and twin sectors of SUSY TH model, regardless of the UV extension.

The one-loop RGEs for MSSM and RHN Yukawa couplings read\footnote{Here, we assume that all the Yukawa matrices, including $y^{i}_n$, are diagonal. The one-loop RGEs for the non-diagonal neutrino Yukawa matrix can be found in~\cite{Hisano:1998fj}. For the general SUSY one-loop RGEs see~\cite{Martin:1997ns}.}
\begin{equation}\label{eq:YukawaRGEs}
\begin{aligned}
    &(4\pi)^2\mu\frac{d y^i_{u}}{d\mu}=y^i_{u}\Big[ \sum_k\big( 3\, (y^k_{u})^2+ (y^k_{n})^2\big)+3 \, (y^i_{u})^2+(y^i_{d})^2-\frac{16}{3} g_3^2-3g_2^2-\frac{13}{15}g_1^2\Big],\\
    &(4\pi)^2\mu\frac{d y^i_{d}}{d\mu}=y^i_{d}\Big[\sum_k\big(3\, (y^k_{d})^2+ (y^k_{l})^2\big)+3\, (y^i_{d})^2+(y^i_{u})^2-\frac{16}{3} g_3^2-3g_2^2-\frac{7}{15}g_1^2\Big],\\
    &(4\pi)^2\mu\frac{d y^i_{l}}{d\mu}=y^i_{l}\Big[\sum_k \big( 3 \, (y^k_{d})^2 +( y^k_{l})^2\big)+3\,(y^i_l)^2+(y^i_n)^2-3g_2^2-\frac{9}{5}g_1^2\Big],\\
    &(4\pi)^2\mu\frac{d y^i_{n}}{d\mu}=y^i_{n}\Big[\sum_k \big( 3 \, (y^k_{u})^2+(y^k_{n} )^2\big)+3\,(y^i_n)^2+(y^i_l)^2-3g_2^2-\frac{3}{5}g_1^2\Big],
\end{aligned}
\end{equation}
where the subscripts $u$, $d$, $l$, and $n$ stand for up-type quark, down-type quark, and charged and neutral lepton Yukawa couplings in the superpotential, respectively, while the upper indices denote the generation number. In scenarios with RHNs, their effective MSSM Yukawas are very small and can be safely neglected. The RGEs for the twin sector are obtained from the MSSM ones by substituting $y^i_j\to\tilde{y}^i_j$. The initial conditions for RGEs \eqref{eq:YukawaRGEs} in the SM sector are set by the observed value and ${\rm tan}\beta$. On the other hand, hard $\mathbb{Z}_2$-breaking in the twin Yukawa sector imposes different initial conditions in TS.

The Yukawa couplings are not asymptotically free for large Yukawa couplings and may develop a Landau pole at some point. The scale at which the RGE solution becomes singular is significantly reduced when Yukawa couplings are enhanced, as their running is proportional predominantly to their own values. We require that the Landau pole appears above the matching scale $\Lambda$, which imposes a theoretical bound on the maximal values of the fermion Yukawas.

The RGEs governing the running of the trilinear scalar couplings at one loop are
\begin{equation}\label{eq:trilinearRGEs}
    \begin{aligned}
        &(4\pi)^2\mu\frac{d a^i_{u}}{d\mu}= a^i_u \Big[\sum_k \big( 3\,(y^k_{u})^2+(y^k_{n})^2\big)+9(y^i_u)^2 + (y_d^i)^2-\frac{16}{3}g_3^2-3g_2^2-\frac{13}{15}g_1^2\Big]\\
        &+2y_u^i\Big[\sum_k\big(3\,a^k_u\, y^k_u+a_n^k y_n^k\big)+\,a^i_d \,y^i_d+\frac{16}{3}g_3^2M_3+3g_2^2M_2+\frac{13}{15}g_1^2 M_1\Big],\\[2mm]
        &(4\pi)^2\mu\frac{d a^i_{d}}{d\mu}= a^i_d \Big[\sum_k \big(3\,(y^k_{d})^2+(y^k_{l})^2\big)+9(y^i_d)^2 + (y_u^i)^2-\frac{16}{3}g_3^2-3g_2^2-\frac{7}{15}g_1^2\Big]\\
        &+2y_d^i\Big[\sum_k\big( 3\,a^k_d\, y^k_d +\, a^k_l \, y^k_l\big)+  a^i_u\,y^i_u+\frac{16}{3}g_3^2M_3+3g_2^2M_2+\frac{7}{15}g_1^2 M_1\Big],\\[2mm]
        &(4\pi)^2\mu\frac{d a^i_{l}}{d\mu}= a^i_l \Big[\,\sum_k \big(3\,(y^k_{d})^2+(y^k_{l})^2\big)+9(y^i_l)^2+(y^i_n)^2-3g_2^2-\frac{9}{5}g_1^2\Big]\\
        &+2y_l^i\Big[\sum_k\big( 3\,a^k_d\, y^k_d +\, a^k_l \, y^k_l\big)+a^i_n\,y^i_n+3g_2^2M_2+\frac{9}{5}g_1^2 M_1\Big],\\[2mm]
    \end{aligned}
\end{equation}
\begin{equation*}
    \begin{aligned}
        &(4\pi)^2\mu\frac{d a^i_{n}}{d\mu}= a^i_n \Big[\,\sum_k \big(3\,(y^k_{u})^2+(y^k_{n})^2\big)+9(y^i_n)^2+(y^i_l)^2-3g_2^2-\frac{3}{5}g_1^2\Big]\\
        &+2y_n^i\Big[\sum_k\big( 3\,a^k_u\, y^k_u +\, a^k_n \, y^k_n\big)+a^i_l\,y^i_l+3g_2^2M_2+\frac{3}{5}g_1^2 M_1\Big],\\[2mm]
    \end{aligned}
\end{equation*}
We do not introduce any direct $\mathbb{Z}_2$-breaking on the trilinear couplings and thus their TS values are fully determined by the initial condition in the SM sector, matching scale $\Lambda$, and RG running in both sectors. For concreteness, in all numerical computations we use the initial condition with the SM trilinear coupling set to zero at the EW scale.

Finally, the RGEs that describe the evolution of Higgs, squarks, and sleptons soft masses read
\begin{equation}
\begin{aligned}\label{eq:muRGEs}
&(4\pi)^2\mu\frac{d\mu^2_{H_u}}{d\mu} = \sum_k\big(3\,X_u^k+X_n^k)-\frac{6}{5}g_1^2 M_1^2-6 g_2^2 M_2^2+\frac{3}{5} g_1^2 S,\\[2mm]
&(4\pi)^2\mu\frac{d\mu^2_{H_d}}{d\mu} = \sum_k\big(3\,X_d^k+X_e^k\big)-\frac{6}{5}g_1^2 M_1^2-6 g_2^2 M_2^2-\frac{3}{5} g_1^2 S,\\[2mm]
&(4\pi)^2\mu\frac{d(\mu_{Q}^i)^2}{d\mu} = X_u^i+X_d^i-\frac{2}{15}g_1^2 M_1^2-6 g_2^2 M_2^2-\frac{32}{3}g_3^2 M_3^2+\frac{1}{5} g_1^2 S,\\[2mm]
&(4\pi)^2\mu\frac{d(\mu^i_{u})^2}{d\mu} = 2 X_u^i-\frac{32}{15}g_1^2M_1^2-\frac{32}{3}g_3^2 M_3^2-\frac{4}{5}g_1^2 S ,\\[2mm]
&(4\pi)^2\mu\frac{d(\mu^i_{d})^2}{d\mu} = 2 X_d^i-\frac{8}{15}g_1^2M_1^2-\frac{32}{3}g_3^2 M_3^2+\frac{2}{5}g_1^2 S,\\[2mm]
&(4\pi)^2\mu\frac{d(\mu^i_{L})^2}{d\mu} = X_n^i+X_e^i-\frac{6}{5}g_1^2M_1^2-6 g_2^2 M_2^2-\frac{3}{5}g_1^2 S,\\[2mm]
&(4\pi)^2\mu\frac{d(\mu^i_{n})^2}{d\mu} = 2 X_n^i,\\[2mm]
&(4\pi)^2\mu\frac{d(\mu^i_{e})^2}{d\mu} = 2 X_e^i-\frac{24}{5}g_1^2M_1^2-\frac{6}{5}g_1^2 S,
\end{aligned}
\end{equation}
where
\begin{align*}
&X^i_u=2\,(y_u^i)^2\Big[\mu_{H_u}^2+(\mu^i_Q)^2+(\mu^i_u)^2\Big]+2\,(a^i_u)^2,\\
&X^i_d=2\,(y_d^i)^2\Big[\mu_{H_d}^2+(\mu^i_Q)^2+(\mu^i_d)^2\Big]+2\,(a^i_d)^2,\\
&X^i_n=2\,(y_n^i)^2\Big[\mu_{H_u}^2+(\mu^i_L)^2+(\mu^i_n)^2\Big]+2\,(a^i_n)^2,\\
&X^i_e=2\,(y_l^i)^2\Big[\mu_{H_d}^2+(\mu^i_L)^2+(\mu^i_e)^2\Big]+2\,(a^i_l)^2,\\
&S=\mu^2_{H_u}-\mu^2_{H_d}+\sum_k\Big[(\mu_Q^k)^2-(\mu_L^k)^2-2(\mu_u^k)^2+(\mu_d^k)^2+(\mu_e^k)^2\Big].
\end{align*}

\subsection{Numerical integration of the RGEs}\label{app:RGEs_res}

\begin{table}[t]
\centering
\begin{tabular}{|lllp{1cm}|}
\hline\hline

\multicolumn{1}{c}{\textbf{Benchmark N\b{o}}} & \multicolumn{1}{c}{\textbf{UV extension N\b{o}}} & \multicolumn{1}{c}{\hspace{4mm}$\boldsymbol{\tilde{y}_l}$\hspace{4mm}} & \multicolumn{1}{c}{\hspace{4mm}$\boldsymbol{\tilde{y}_n}$\hspace{4mm}} \\[1mm] \hline\hline

\multicolumn{1}{c}{\textbf{1.}}  & \multicolumn{1}{c}{\textbf{2)}}  & \multicolumn{1}{c}{$1.5$}   & \multicolumn{1}{c}{--} \\[1mm] \hline \hline

\multicolumn{1}{c}{\textbf{2.}}  & \multicolumn{1}{c}{\textbf{3)}}  & \multicolumn{1}{c}{$0.9$}   & \multicolumn{1}{c}{0.9} \\[1mm] \hline \hline

\multicolumn{1}{c}{\textbf{3.}}  & \multicolumn{1}{c}{\textbf{4)}}  & \multicolumn{1}{c}{$0.9$}   & \multicolumn{1}{c}{0.7} \\[1mm] \hline \hline

\end{tabular}
\caption{Three benchmark points for which in the UV extensions \textbf{2)}-\textbf{4)} summarized in table \ref{tab:TQ} predict FOPT and SNR. In the first benchmark FOPT requires $\tan\beta\leq 1.5$, while for the remaining two $\tan\beta\leq 3$ is sufficient.
}
\label{tab:Bench}
\end{table}

In the second part of this appendix, we numerically integrate the one-loop RGEs \eqref{eq:YukawaRGEs}-\eqref{eq:muRGEs} in the MSSM and TS. First, we compute the running of gauge and Yukawa couplings as well as SUSY soft terms in the MSSM sector, starting from the EW scale at which they are fixed, up to the scale of $\mathbb{Z}_2$-breaking $\Lambda$. At this scale, gauge couplings and soft SUSY terms are matched with their TS analogues. The RGEs for gauge couplings do not depend on any other couplings/soft terms at one loop and therefore the TS solution is identical to the MSSM one. The explicit $\mathbb{Z}_2$-breaking in the Yukawa couplings is implemented by imposing their values in TS at the scale $\mu=f_0$ and solving the subsystem \eqref{eq:YukawaRGEs} that determines their running up to $\Lambda$ scale. 
 Finally, the running of the soft SUSY terms in TS is found with the help of eq. \eqref{eq:trilinearRGEs} and \eqref{eq:muRGEs}. The twin SUSY-breaking parameters run from $\mu=\Lambda$, at which they match the MSSM soft terms, down to TH scale $f_0$, at which they are used to estimate the minimal tuning \eqref{eq:FT} and the splitting in the soft masses of charged sleptons
\begin{equation}\label{eq:mu_spl}
\Delta\mu^2_{L,\;e}\equiv\hat{\mu}^2_{L,\;e}-\mu^2_{L,\;e}.
\end{equation}
We will discuss how these quantities depend on the matching scale $\Lambda$ in the UV extensions considered in this paper.

\begin{figure}[t]
    \centering
    \subcaptionbox{Benchmark \textbf{1} (green)}{
    \includegraphics[width=0.45\textwidth]{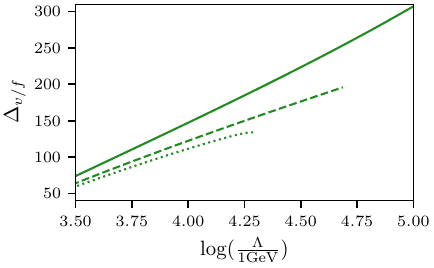}}\hspace{5mm}
    \subcaptionbox{Benchmarks \textbf{2} (blue) and \textbf{3} (red)}{
    \includegraphics[width=0.45\textwidth]{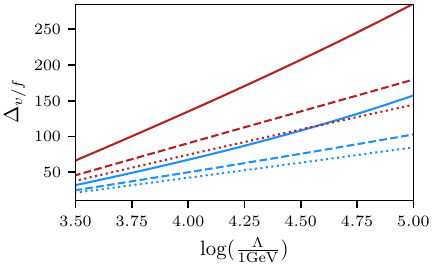}}\hspace{5mm}
        \caption{The minimal tuning as a function of $\mathbb{Z}_2$-breaking scale $\Lambda$ for three benchmarks from table \ref{tab:Bench}. Solid, dashed and dotted lines depict $\tan\beta$ equal to $1$, $2$ and $3$ respectively. Some lines end before $\Lambda=10^5$ GeV since the down-type quark Yukawas hit the Landau pole. For LH and RH charged sleptons we used the same initial condition $(200\;\text{GeV})^2$ at a scale $\mu=100$ GeV, on the SM site, while for RH neutrinos we fix $\mu^2_n=(10\;\text{GeV})^2$. The soft squark masses in the SM sector are set to $\mu_Q^2=\mu_u^2=\mu_d^2=(2 \text{ TeV})^2$ at $\mu=100$ GeV.}
    \label{fig:SleptonTuning}
\end{figure}

In figure~\ref{fig:SleptonTuning} we plotted the minimal tuning versus $\Lambda$ for three representative benchmarks summarized in table \ref{tab:Bench} that feature both the SNR and FOPT. Notably, for all benchmarks minimal tuning grows with the matching scale. Naturally, this is a consequence of longer RG running -- for larger $\Lambda$ values, the $\mathbb{Z}_2$-breaking in fermion Yukawas induces bigger splitting between the soft Higgs masses $\Delta m^2$ to which the minimal FT is proportional (see eqs.~\eqref{eq:FT_v/f} and \eqref{eq:FT}). On the other hand, fixing $\Lambda$ close to the TH scale $f_0$ could inflict unexpected interference between the mechanism responsible for $\mathbb{Z}_2$ breaking and the TH UV completion dictated by the little hierarchy problem. Moreover, such model can seriously affect the thermal Higgs phase evolution, rendering our main results inadequate. Hence, we do not consider such possibility, and cut the plots at $10^{3.5}$ GeV.

The plots in figure \ref{fig:SleptonTuning} suggest that in the minimal UV extension \textbf{2)} with only sleptons being light, there is no room for pushing the new physics scale $\Lambda$ beyond $10^4$ GeV since the minimal FT is already high at that value. Nevertheless, with UV extensions that contain twin RHNs, one can consider larger $\Lambda$ values and still obtain models that are more natural than the ordinary MSSM extension of the SM. We also observe that for each benchmark, the minimal tuning is larger for smaller $\tan\beta$ values, which indicates that in the class of SUSY TH extensions discussed in this paper FT is dominated by the splitting in down-type Higgs soft mass.

\begin{figure}[t]
    \centering
    \subcaptionbox{Left-handed sleptons.}{
    \includegraphics[width=0.45\textwidth]{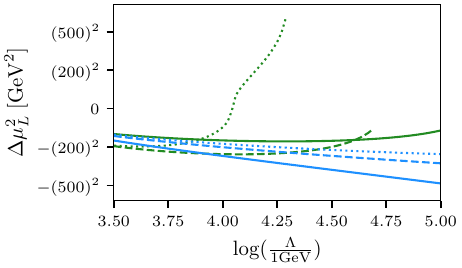}}\hspace{5mm}
    \subcaptionbox{Right-handed charged sleptons.}{
    \includegraphics[width=0.45\textwidth]{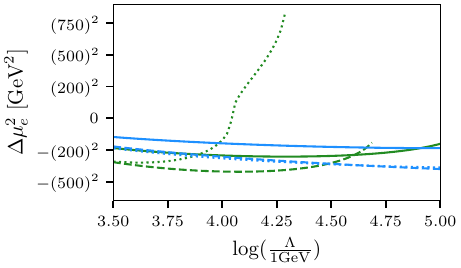}}\hspace{5mm}
        \caption{The soft slepton mass splitting \eqref{eq:mu_spl} at TH scale $\mu=1$ TeV. The results obtained for benchmarks no. \textbf{1} and \textbf{2} from table \ref{tab:Bench} are plotted with green and blue lines. The initial values of the soft masses in the SM sector and $\tan\beta$ are set as in figure \ref{fig:SleptonTuning}.}
    \label{fig:SleptonMasses}
\end{figure}

 Figure~\ref{fig:SleptonMasses} shows the soft mass splitting $\Delta\mu^2_{L,\,e}$ versus $\Lambda$ expected in UV extensions \textbf{2)} and \textbf{3)}. Clearly, splitting in the soft masses of charged sleptons of the order $-(300\;\text{GeV})^2$ at $\Lambda=10^4\;\text{GeV}$ is a generic result for these two scenarios. Negative $\Delta\mu^2_{L,e}$ is a direct consequence of enhancing twin lepton Yukawas. It follows from eq. \eqref{eq:muRGEs} that for $y_{l}\ll \tilde{y}_{l}$ slepton soft terms gain much less in the MSSM up-scale evolution than they lose when they run down-scale in the TS. In the main part of this work, the splitting in the soft terms was not computed directly, but instead some specific values were picked that were optimal for phase transition and did not render any physical slepton mass negative. This approach is methodologically fine since in the MSSM sector there is some freedom in choosing the exact values of the soft slepton terms. Here, we explicitly verify that twin soft masses used for the scans \ref{fig:SNRmu1}, \ref{fig:SNRmu4} and \ref{fig:fScan} can be obtained for the natural slepton soft masses on the MSSM side that are still consistent with the lower bound set by the LEP at $90$ GeV~\cite{particle2022,ALEPH:2001oot,DELPHI:2003uqw,OPAL:2003nhx,L3:2003fyi}.

\section{Low-scale UV completion for $\mathbb{Z}_2$-breaking Yukawas}\label{app:Vquarks}

As can be seen from Appendix.~\ref{app:RGEs_res},
in order to avoid a fine-tuned EW scale, the large $\mathbb{Z}_2$-breaking Yukawas call for a low $\Lambda \lesssim 10^4$ GeV below which the soft Higgs masses receive $\mathbb{Z}_2$-breaking quantum corrections. In this appendix we describe a perturbative UV completion with low $\Lambda$.

As an illustration, we discuss the up Yukawa coupling. Application to other Yukawa couplings is straightforward. We introduce a Dirac fermion $U$ and $\bar{U}$ and their twin partners. The UV-complete superpotential is
\begin{equation}
    W = y H_u q \bar{U} + M U (\bar{U} + \epsilon \bar{u}) + y H_u' q' \bar{U}' + M U' (\bar{U}' + \epsilon' \bar{u}'),
\end{equation}
where we introduce soft $\mathbb{Z}_2$ breaking such that $\epsilon' \neq \epsilon$.
Above the scale $M$, the running of the soft Higgs mass is $\mathbb{Z}_2$ symmetric.
Below the scale $M$, we may integrate out the Dirac fermions and obtain an effective superpotential
\begin{equation}
    W_{\rm eff} = y \epsilon H_u q \bar{u}+  y\epsilon' H_u' q' \bar{u}'.
\end{equation}
The SM up Yukawa is $y \epsilon$ and the twin up Yukawa is $y \epsilon'$. By taking $\epsilon' \gg \epsilon$, we obtain $y_u' \gg y_u$. Below the scale $M$, the running of the soft Higgs mass becomes $\mathbb{Z}_2$ breaking. We may identify $M$ with $\Lambda$.

\section{UV completion of neutrino Yukawa couplings}
\label{app:UV_RHN}
In this appendix we present a UV completion of the neutrino Yukawa coupling $y_n$. To be concrete, we discuss the D-term Twin Higgs model in~\cite{Badziak:2017wxn}, where the SM Higgs and Twin Higgs are embedded into fundamental representation of $SU(2)_X$ and $SU(2)_X'$, respectively. $SU(2)_X\times SU(2)_X'$ is broken down into $SU(2)_D$ whose D-term potential is responsible for the $SU(4)$ symmetric quartic coupling in the Higgs potential. 

The $SU(2)_X\times SU(2)_L\times U(1)_Y\times U(1)_{B-L}$ charges of the fields relevant for the neutrino mass are shown in table~\ref{tab:charge_nu}. The Twin sector has the mirror copy of them. $\phi$ and $\bar{\phi}$ are $B-L$ breaking field, and $S$ and $\bar{S}$ are $SU(2)_D$ breaking fields.
The superpotential term of the neutrino sector is
\begin{equation}
    W = y_n H L \bar{N} + m \bar{N}\psi + y_S S \psi \bar{\chi} + y_{\bar{S}} \bar{S} \psi \bar{\chi} + \lambda_N \phi \bar{N}\bar{N} + \lambda_\psi \bar{\phi} \psi \psi + \lambda_\chi \phi \bar{\chi} \bar{\chi},
\end{equation}
where we suppressed the generation and $SU(2)_X$ indices. We assume $\mathbb{Z}_2$ symmetry in the couplings and masses in this superpotential.

\begin{table}
    \centering
    \begin{tabular}{c|cccc}
         & $SU(2)_X$  & $SU(2)_L$ & $U(1)_Y$ & $U(1)_{B-L}$   \\ \hline
     $H$    & {\bf 2} & {\bf 2}  & $1/2$  & $0$ \\
     $L_{1,2,3}$  & {\bf 1}  & {\bf 2}  & $-1/2$  & $-1$ \\
     $\bar{N}_{1,2,3}$ & {\bf 2}  & {\bf 1}  & $0$  & $1$ \\
     $\psi_{1,2,3}$    & {\bf 2}  & {\bf 1}  & $0$  & $-1$ \\ 
     $\bar{\chi}_{1,2,3}$ & {\bf 1} & {\bf 1} & $0$ & $1$ \\
     $\phi$ & {\bf 1} & {\bf 1} & $0$ & $-2$ \\
     $\bar{\phi}$ & {\bf 1} & {\bf 1} & $0$ & $2$ \\
     $S$ & {\bf 2} & {\bf 1} & $0$ & $0$ \\ 
     $\bar{S}$ & {\bf 2} & {\bf 1} & $0$ & $0$
    \end{tabular}
    \caption{Charges of the fields relevant for the neutrino sector in the generalised SUSY $D$-term TH model.}
    \label{tab:charge_nu}
\end{table}

In the Twin sector, $\phi'$ and $\bar{\phi}'$ do not obtain VEVs. By taking $m \ll y_n v_B$  and $y_n= \mathcal{O}(1)$, the neutrino Yukawa around the eletroweak phase transition temperature can be indeed $\mathcal{O}(1)$. In the SM sector, $\phi$ and $\bar{\phi}$ obtain VEVs. In order to suppress the neutrino Yukawa, $\bar{N}$ must obtain a mass much above the electroweak scale. This can be done by large $\lambda_N \phi$ or large $m$ with small $\lambda_\psi \bar{\phi}$. 
Non-zero neutrino masses may be generated by the seesaw or inverse-seesaw mechanism, depending on the hierarchy between the $B-L$ preserving mass terms and the $B-L$ breaking mass terms.

\medskip
\bibliography{sources}

\end{document}